\documentclass{JHEP3}

\preprint{}

\keywords{{Large Extra Dimensions}, {Classical Theories of Gravity}}
\usepackage{graphicx}
\usepackage{latexsym}
\usepackage{amsmath}
\usepackage{amsfonts}
\usepackage{amssymb}
\newcommand{\be}{\begin{equation}}
\newcommand{\ee}{\end{equation}}
\newcommand{\ben}{\begin{eqnarray}}
\newcommand{\een}{\end{eqnarray}}
\newcommand{\bes}{\begin{subequations}}
\newcommand{\ees}{\end{subequations}}

\newcommand{\bb}{\bibitem}

\title{Gravity localization on thick branes: a numerical approach}

\author{D. Bazeia\\Departamento de F\'\i sica, Universidade Federal da Para\'\i ba, Jo\~ao Pessoa PB, Brazil\\
Email: bazeia@fisica.ufpb.br}

\author{A. R. Gomes\\Departamento de Ci\^encias Exatas, Centro Federal de Educa\c c\~ao Tecnol\'ogica do Maranh\~ao, S\~ao Lu\'\i s MA, Brazil\\
E-mail: {argomes@pq.cnpq.br}}

\author{L. Losano\\Departamento de F\'\i sica, Universidade Federal da Para\'\i ba, Jo\~ao Pessoa PB, Brazil\\
Email: losano@fisica.ufpb.br}

\abstract{We introduce a numerical procedure to investigate the spectrum of massive modes and its contribution for gravity localization
on thick branes. After considering a model with an analytically known Schroedinger potential, we present the method and discuss its
applicability. With this procedure we can study several models even when the Schroedinger potential is not known analytically. We discuss both the occurrence of localization of gravity and the correction to the Newtonian potential given by the massive modes.
\vspace{2cm}}
\begin{document}
\section{Introduction}

In this work we deal with (4+1)-dimensional actions describing a scalar field coupled to gravity which leads to brane solutions with a single extra dimension \cite{RS,RS1,df,csaki,Brane,brane1}. As one knows, after an {\it Ansatz} is given for the metric, minimization of the action leads to a
system of second-order differential equations for the field and the metric parameter related to the warp factor. With a convenient choice
of the potential one can obtain a system of first-order equations, which helps to find  analytic solutions for the metric and field.

In this scenario, fluctuation around the solutions for the metric and scalar field can be decoupled in the transverse traceless gauge.
After dropping the (3+1)-dimensional plane wave components, considered to satisfy a Klein-Gordon equation, the extra-dimension component
of the metric fluctuations is reduced to a Schroedinger-like equation with an infinite number of solutions. Each solution is assigned to
a massive mode characterized by the mass-shell condition imposed after the separation of the contribution of the four standard dimensions
from the fifth or extra dimension.

In order to obtain the Schroedinger-like equation one usually goes to a conformal variable related to the extra dimension. However,
it occurs that such a transformation is not always possible to be obtained analytically, and this may lead to a Schroedinger potential which
is not known analytically for most of problems considered in the literature. Also, even when one knows the Schroedinger potential analytically,
the whole spectrum of massive modes is hard to be obtained and normalized properly in order to study the main correction to the Newtonian potential.

In the literature, Ref. \cite{csaki} has already shown that any of the following two conditions is sufficient to prove the occurrence
of gravity localization, namely one has to: i) consider the zero-mode, its normalization and asymptotic behaviour or ii) analyze the asymptotic
behaviour of the Schroedinger potential. For instance, both procedures were used in Ref. \cite{bfg} to confirm gravity localization in a specific
class of models.

However, besides confirming the occurrence of gravity localization in a model, it is of physical significance to estimate the first-order
corrections to the Newtonian potential. This can be related to the limit imposed by experiments on gravity at short distances \cite{exper}.
In this direction, Ref. \cite{csaki} also presented an asymptotic analysis for the massive modes in a class of potentials that falls off as
$\alpha(\alpha+1)/z^2$ far from the brane; as a result, a correction of the order of $1/R^{2\alpha-1}$ is found. The issue is that corrections
to the Newtonian potential for analytical Schroedinger-like potentials can in principle be found in a similar way. However, it is not rare to
face problems in which the Schroedinger-like potentials are only known numerically, and in such cases the above procedure cannot be used anymore.
This poses the question on how one can generate and normalize properly the massive modes in order to obtain the corrections to the Newtonian potential.
The issue has motivated us to investigate the problem, and to propose the present procedure. Our numerical study shows that in order for
the spectrum of massive modes to be obtained properly one must apply the normalization procedure with great care. A wrong choice of the plane
wave normalization can lead to a misunderstanding of the pattern of the distribution of the massive modes and to a wrong prediction of the corrections.

In the present work, in the next Sec.~{\ref{sec2}} we deal with the equations one needs to build a Minkowski brane and to study the corresponding
fluctuations
in the metric and scalar fields. In Sec.~\ref{volcano} we review the analytic model of Ref.~\cite{Brane} and there we introduce the
numerical method for the corrections of the Newtonian potential. A comparison of our result and the asymptotic analysis of \cite{csaki}
is done in order to evaluate the precision of the method. We then discuss in Sec. \ref{secanalit} some models with analytically
known Schroedinger potential where we apply the known asymptotic analysis method, and in Sec. \ref{secnumeric} we study  models
with numerically known Schroedinger potential, where we apply our numerical method. A comparison between the results coming from
the numerical method and the asymptotic analysis is done. Also a comparison between the results coming from the numerical Numerov
method and a simpler Euler method also available for non-analytic potentials is done. We end this work in Sec.~{\ref{secconcl}}, where
we include some conclusions and perspectives of future
investigations. A short review of the Numerov method for determining the massive modes is presented in an Appendix.

\section{The brane equations and fluctuations}
\label{sec2}

We start with
\ben
\label{action}
S=\int d^4x\,dy\sqrt{|g|}\Bigl[-\frac14 R+
\frac12\partial_a\phi\partial^a\phi-
V(\phi)\Bigr]\,,
\een
where $g=\det(g_{ab})$ and the metric
\ben
ds^2=g_{ab}dx^adx^b=e^{2A}\eta_{\mu\nu}dx^{\mu}dx^{\nu}-dy^2
\een
describes a background with 4-dimensional Poincare symmetry with $y$ as the extra dimension.
Here $a,b=0,1,2,3,4,$ and $e^{2A}$ is the warp factor. We suppose that
the scalar field and the warp factor only depend on the extra coordinate $y$.

The action given by Eq.(\ref{action}) leads to the following equations for the scalar field $\phi(y)$ and the function
$A(y)$ from the warp factor:
\bes\ben\label{eom2}
\phi^{\prime\prime}+4A^\prime\phi^\prime&=& \frac{dV(\phi)}{d\phi}
\\
A^{\prime\prime}&=&-\frac23\,\phi^{\prime2}
\\
A^{\prime2}&=&\frac16\phi^{\prime2}-\frac13 V(\phi)
\een\ees
where the prime is used to represent derivative with respect to $y$.

The potential is supposed to have the form \ben\label{gpot}
V(\phi)=\frac18 \left(\frac{dW}{d\phi}\right)^2 -\frac13 W^2 \een
where $W(\phi)$ is in principle an arbitrary function of the field
$\phi$ -- in the supersymmetric context $W$ is named
superpotential. The particular relation between $V$ and $W$ in
\eqref{gpot} leads to a description in terms of a set of
first-order differential equations, which are given by
\cite{df,Brane} \bes\label{eom1} \ben
\phi^{\prime}&=&\frac12\,\frac{\partial
W}{\partial\phi}\label{e1a}
\\
A^\prime&=&-\frac13\,W\label{Aprime}
\een\ees

We consider now the effective 4-dimensional gravitational fluctuations in the conformally flat background
discussed previously, as well as the fluctuation
of scalar fields around solutions of the first-order equations \eqref{eom1}; that is, we write:
\ben
ds^2=e^{2A(y)}(\eta_{\mu\nu}+\epsilon h_{\mu\nu})dx^\mu dx^\nu-dy^2
\een and we set $\phi\rightarrow\phi+\epsilon\tilde{\phi}$  where $h_{\mu\nu}=h_{\mu\nu}(x,y)$ and
$\tilde\phi=\tilde\phi(x,y)$ represent the fluctuations and $\epsilon$ is a small parameter. On the transverse traceless gauge
the metric perturbation separates from the scalars \cite{df}, leading to
\ben\label{h} {\bar h}_{\mu\nu}^{\prime\prime}+4\,A^{\prime} \,{\bar
h}_{\mu\nu}^{\prime}=e^{-2A}\,\Box\,{\bar h}_{\mu\nu},
\een
with $\Box$ being the $(3+1)$-dimensional d'Alembertian.

This equation can be decoupled by separating the 4-dimensional plane wave perturbations from the extra dimension contribution.
We introduce a new variable $z$ that turns the metric into a conformal one. This changes the equation for the extra dimension
 contribution of the metric perturbations in a Schroedinger-like form, where no single derivative terms are present. The new conformal
 coordinate $z$ is defined by
\ben\label{dz}
dz=e^{-A(y)}dy.
\een
The separation of variables is taken as
\ben
{\bar h}_{\mu\nu}(x,z)=e^{ip\cdot x}e^{-\frac{3}{2}A(z)}\psi_{\mu\nu}(z),
\een
which turns Eq.~(\ref{h}) into a Klein-Gordon equation for the 4-dimensional components of the transverse-traceless
${\bar h}_{\mu\nu}$, with the remaining Schroedinger-like equation
\ben\label{se}
-\frac{d^2\psi_m(z)}{dz^2}+V_{sch}(z)\,\psi_m(z)=m^2\,\psi_m(z)
\een
where we have dropped the $\mu\nu$ indices from the wavefunctions, now labeled by the corresponding energy $m^2$. Here
the Schroedinger-like potential is given by
\ben\label{Uz}
V_{sch}(z)=\frac32\,A^{\prime\prime}(z)+\frac94\,A^{\prime2}(z)
\een
and this ends our revision of the standard braneworld scenario.

\section{An analytic volcano potential}
\label{volcano}

Before going into the issue concerning our numerical procedure, let us first consider the construction  of a simple analytic potential
characteristic of gravity localization. According to \cite{csaki}, for potentials constructed with the procedure of Eq.~(\ref{Uz}) that go
to infinity as $1/z^2$ have this property. Also, in order to have a volcano structure we need the two components of the rhs of \eqref{Uz} with
opposite signs. A simple choice for $A(z)$ is
\ben
\label{Az}
A(z)=-\ln(1+z^2),
\een
which is depicted in Fig.~\ref{figUz}. In this case Eq.~(\ref{Uz}) gives the related Schroedinger-like potential
\ben
V_{sch}=-\frac3{(1+z^2)}+\frac{15z^2}{(1+z^2)^2}.
\een
which is also shown in Fig.~\ref{figUz}

\FIGURE{\includegraphics[{angle=0,width=7cm}]{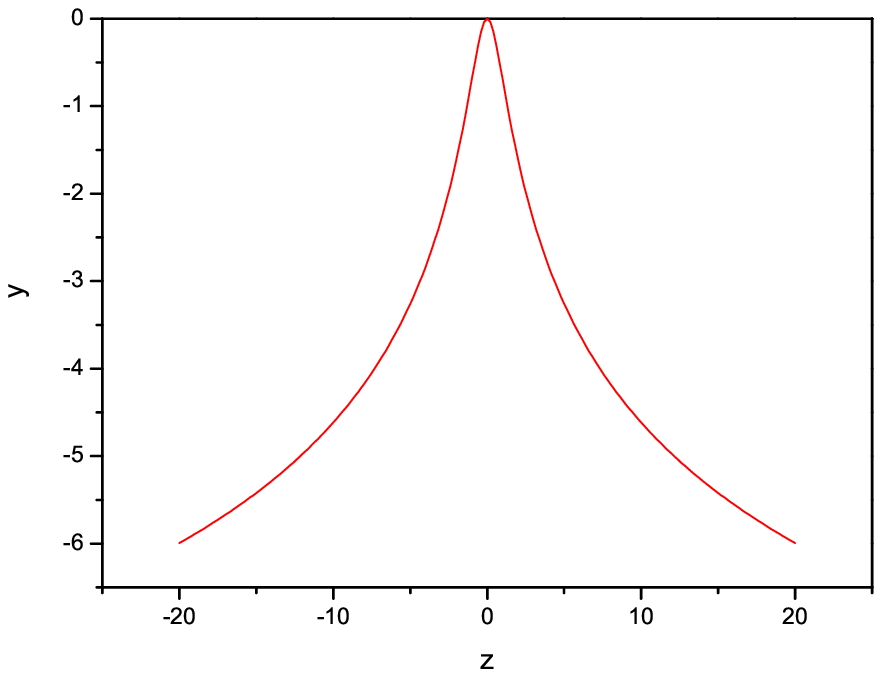}
\includegraphics[{angle=0,width=7cm}]{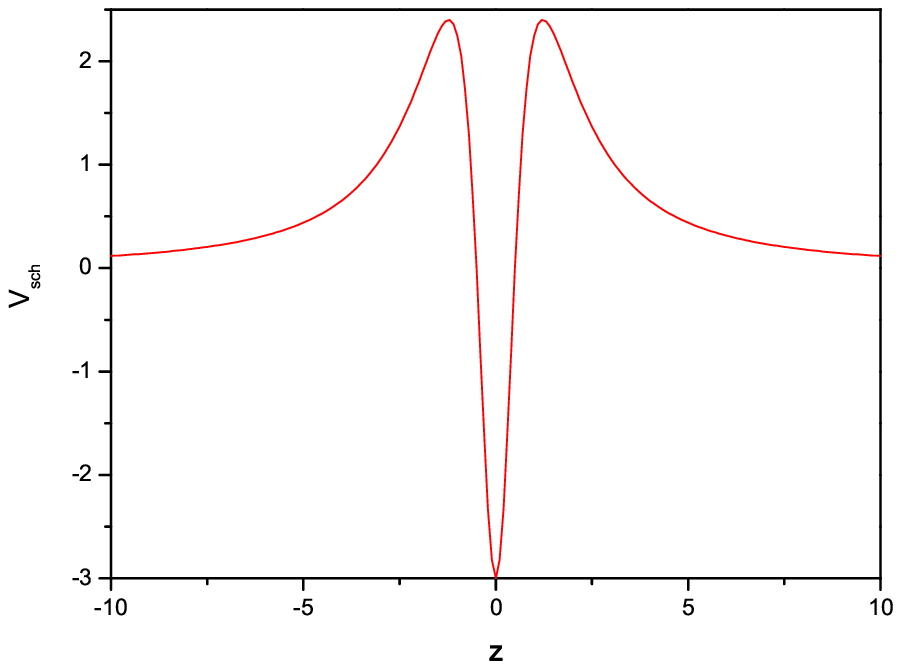}
\caption{Plot of $A(z)$ (left) and the Schroedinger-like potential $V_{sch}(z)$ (right).}
\label{figUz}
}

The characteristic volcano shape of the potential leads to the thick brane solution, so it contributes to smooth the potential considered
in the original Randall-Sundrum \cite{RS} model which describes the thin brane solution. The analytic structure of $V_{sch}(z)$ eases the
 investigation of gravity localization. In particular, the zero-mode is determined analitically as
\ben
\label{zeromode}
\psi_0(z)=\frac {N_0}{(1+z^2)^{3/2}},
\een
where $N_0$ is a normalization constant.

\subsection{Reconstructing the model}

Before turning to the gravity localization issue, let us first
reverse the investigation and work on the reconstruction of the
model. This means to obtain the functions $A(y)$, $z(y)$,
$\phi(y)$ and $V(\phi)$ once $A(z)$ is given. We hope that this
reconstruction process will give us insights on the importance of
$V_{sch}(z)$ in the far-reaching region for gravity localization.

We first consider Eq.~\eqref{dz} and write
\ben
\label{express_zy}
z=\int e^{-A(y)} dy.
\een
Suppose we can write analytically $z(y)$ in order to change the representation as $A(z)=A(z(y))=A(y)$. In this way we can
 invert Eq.~(\ref{dz}) and write, in the present case
\ben
y=\int e^{A(z)} dz= \int e^{-\ln{(1+z^2)}}dz = \tan^{-1}(z).
\een
We invert this relation to write
\ben
z=\tan(y)
\een
which we plot in Fig.~\ref{figAy}. We use it in Eq.~(\ref{Az}) to get
\ben
\label{Ay}
A(y)=-\ln{(1+\tan^2(y))}.
\een
which shows -- see Fig.~\ref{figAy} -- that the $y$ variable is allowed to vary continuously only when spanning finite intervals,
 characterizing an effective compactification of the extra dimension in this representation.

\FIGURE{\includegraphics[{angle=0,width=7cm}]{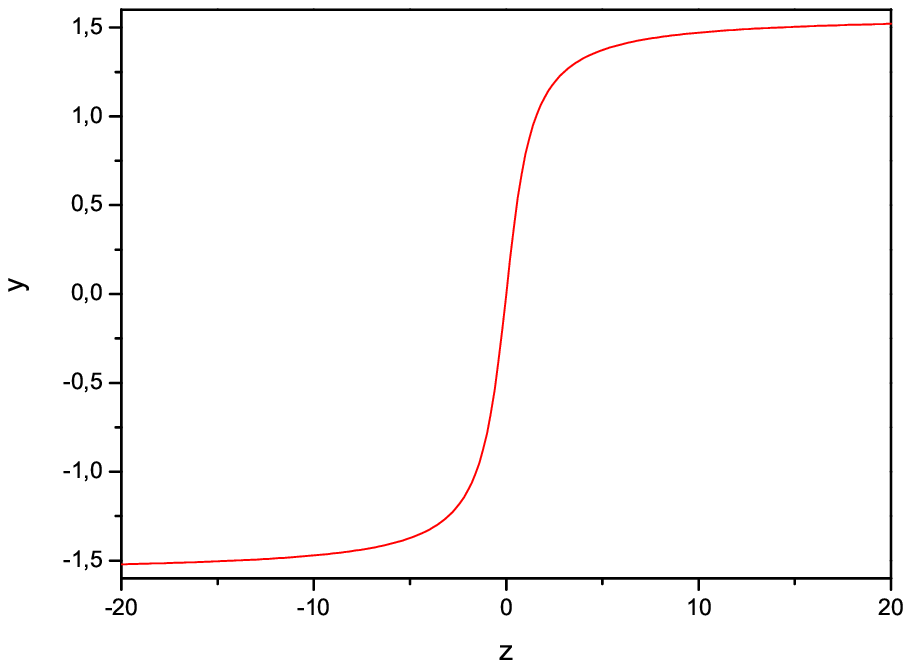}
\includegraphics[{angle=0,width=7cm}]{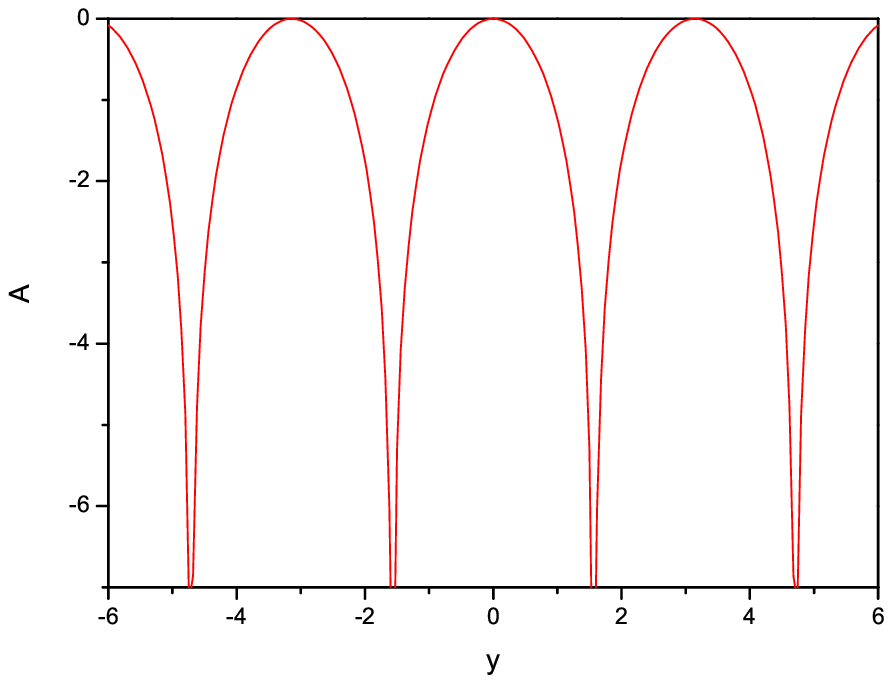}
\caption{Plots of the functions $y(z)$ (left) and $A(y)$ (right). Note the effective compactification on the extra dimension y.}
\label{figAy}}

The warp factor is shown in Fig. \ref{figwarp}. There we see that it goes to zero far from the brane, as usual, but here this happens in
 a finite distance $y^*$ from the brane, achieving the characteristic $AdS_5$ space in this region.

\FIGURE{\includegraphics[{angle=0,width=7cm}]{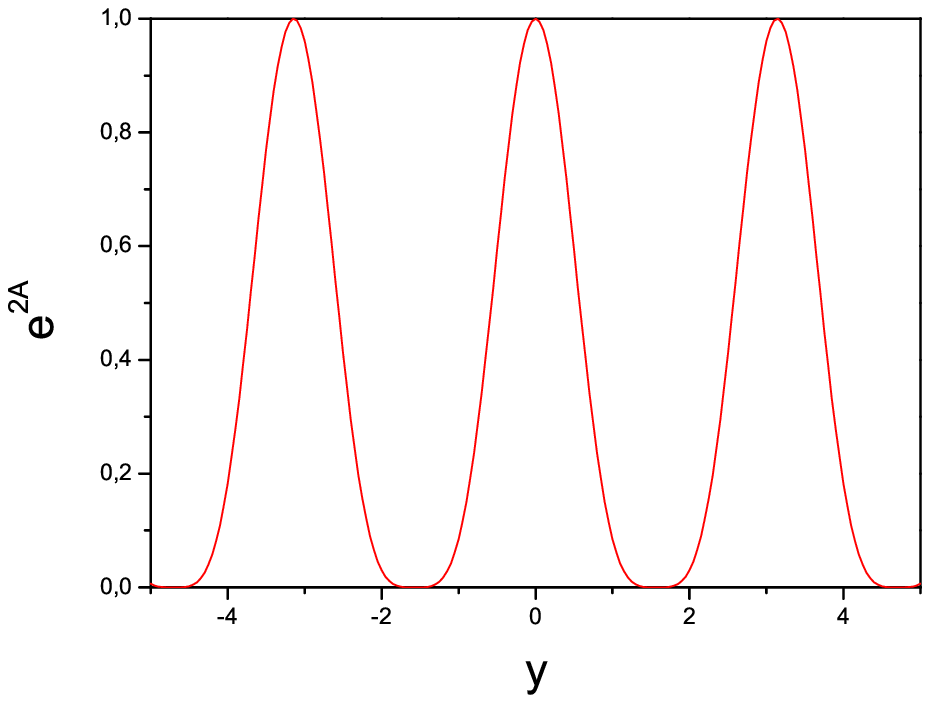}
\includegraphics[{angle=0,width=7cm}]{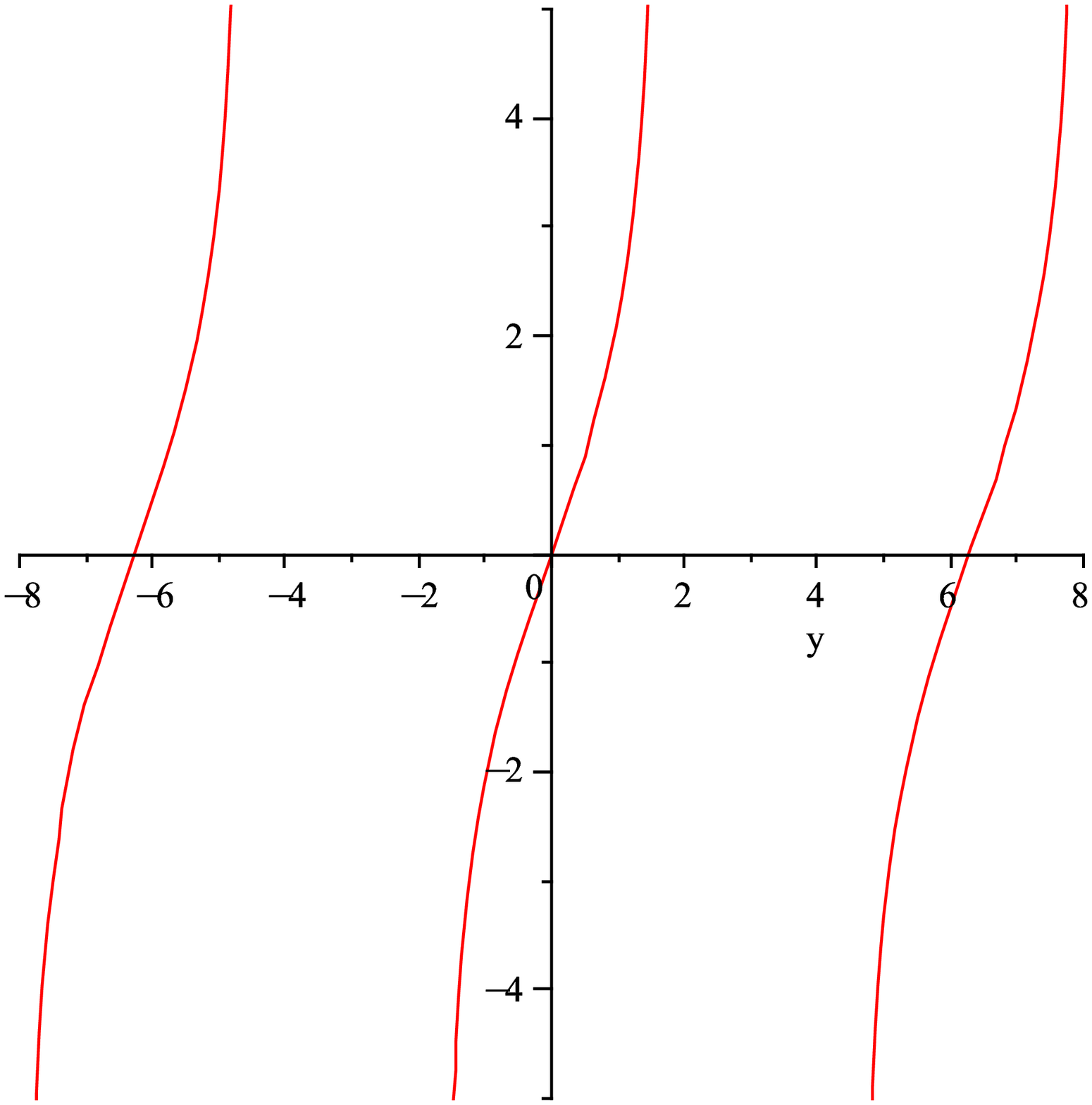}
\caption{Plots of the warp factor $e^{2A(y)}$ (left) and scalar field $\phi(y)$ (right). Note the effective compactification on the extra
 dimension y.}
\label{figwarp}}

The formation of a natural barrier on the extra dimension $y$ must also be seen in the solution of the scalar field $\phi(y)$. In order to
 recover this, we turn attention to Eqs.~\eqref{eom1}. Differentiating Eq.~(\ref{Ay}) with respect to $y$ and substituting in Eq. (\ref{Aprime})
  gives $W(\phi)=6\tan(y)$. We can differentiate this to get $W'(\phi)\phi'(y)=6\sec^2{y}$. We now use Eq.~(\ref{e1a}) to obtain
\ben
\label{phiy}
\phi(y)=\sqrt{3}\ln({\sec y + \tan y}).
\een
In Fig.~\ref{figwarp} one sees that there are finite intervals of $y$ in which the function $\phi(y)$ is not defined, and the brane has a
diffuse structure. This can be better observed if we turn back to the $z$ coordinate; we write
\ben
\phi(z)=\sqrt(3)\ln{(\sqrt{1+z^2}+z)},
\een
which is plotted in Fig.~\ref{figphiz}, showing the kinklike profile corresponding to a difuse wall, as it appears in the vacuumless
potential investigated in \cite{V}.

\FIGURE{\includegraphics[{angle=0,width=7cm}]{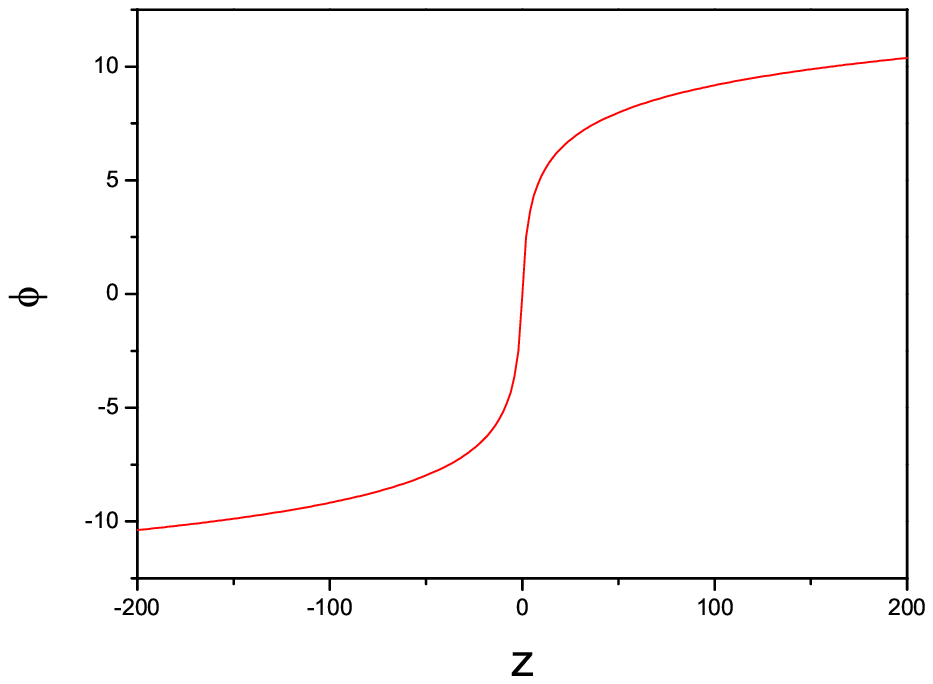}
\includegraphics[{angle=0,width=7cm}]{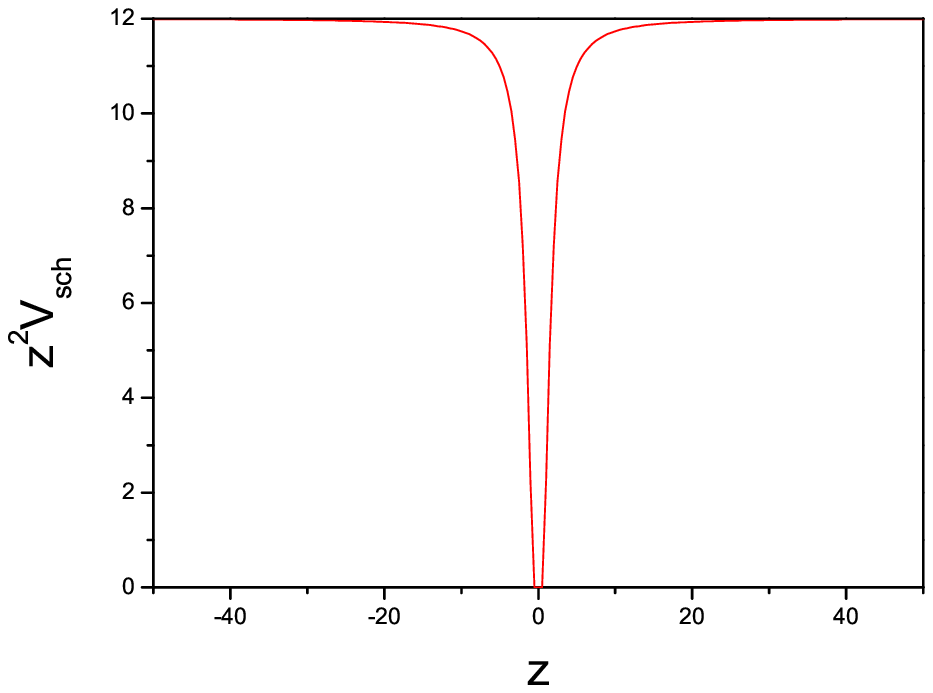}
\caption{Plots of $\phi(z)$ (left), showing resemblance with a kink struture, but without an asymptotic value for $\phi(z),$ and of
$z^2V_{sch}(z)$ (right), showing the asymptotic regime for the Schroedinger-like potential.}
\label{figphiz}}

In this case the superpotential can also be analytically determined in the following way. An ordinary differential equation of first
 order for $W(\phi)$ is obtained after using Eqs.~{\eqref{eom1}} and derivatives of $A(y)$ in Eq.~{\eqref{Ay}} and of $\phi(y)$ in Eq.~{\eqref{phiy}.
The procedure leads to $W'(\phi)+1/\sqrt3W(\phi)=2\sqrt3\exp{\phi/\sqrt3}$. We solve this equation to obtain
\ben
W(\phi)=3e^{2/3\phi\sqrt3}-A
\een
with $A$ being a free parameter. The potential is then given by
\ben
V(\phi)=-\frac32e^{\frac43\phi\sqrt3}+2Ae^{\frac23\phi\sqrt3}-\frac13A^2
\een

The analytic model is very nice, and it leads us to go forward or backward very nicely.

\subsection{Massive modes}

Now we turn back to the Schrodinger-like equation with the aim to
study the massive modes. As one knows, in the present scenario it
is important to find the full spectrum of massive modes. The issue
is of greater importance within the context of local localization
of gravity \cite{KR}. Although it is not always possible to find
analytical expressions for the massive modes \cite{ref_analit},
the analytical form of the potential $V_{sch}$ helps us very much
to control the numerical investigation. Indeed, a simple
Runge-Kutta routine can be used with the boundary conditions
$\psi_m(0)=1$ and $\psi_m'(0)=0$. The first condition is
arbitrarily fixed, and will be adjusted with the normalization
procedure. As one knows, the conformal $z-$variable leads to a
probabilistic interpretation for $|\psi_m(z)|^2$ and one must
impose the condition \ben \label{normproc}
\int_{-\infty}^{\infty}|\psi_m(z)|^2 dz =1. \een

\FIGURE{\hspace{2cm}\includegraphics[{angle=0,width=7cm}]{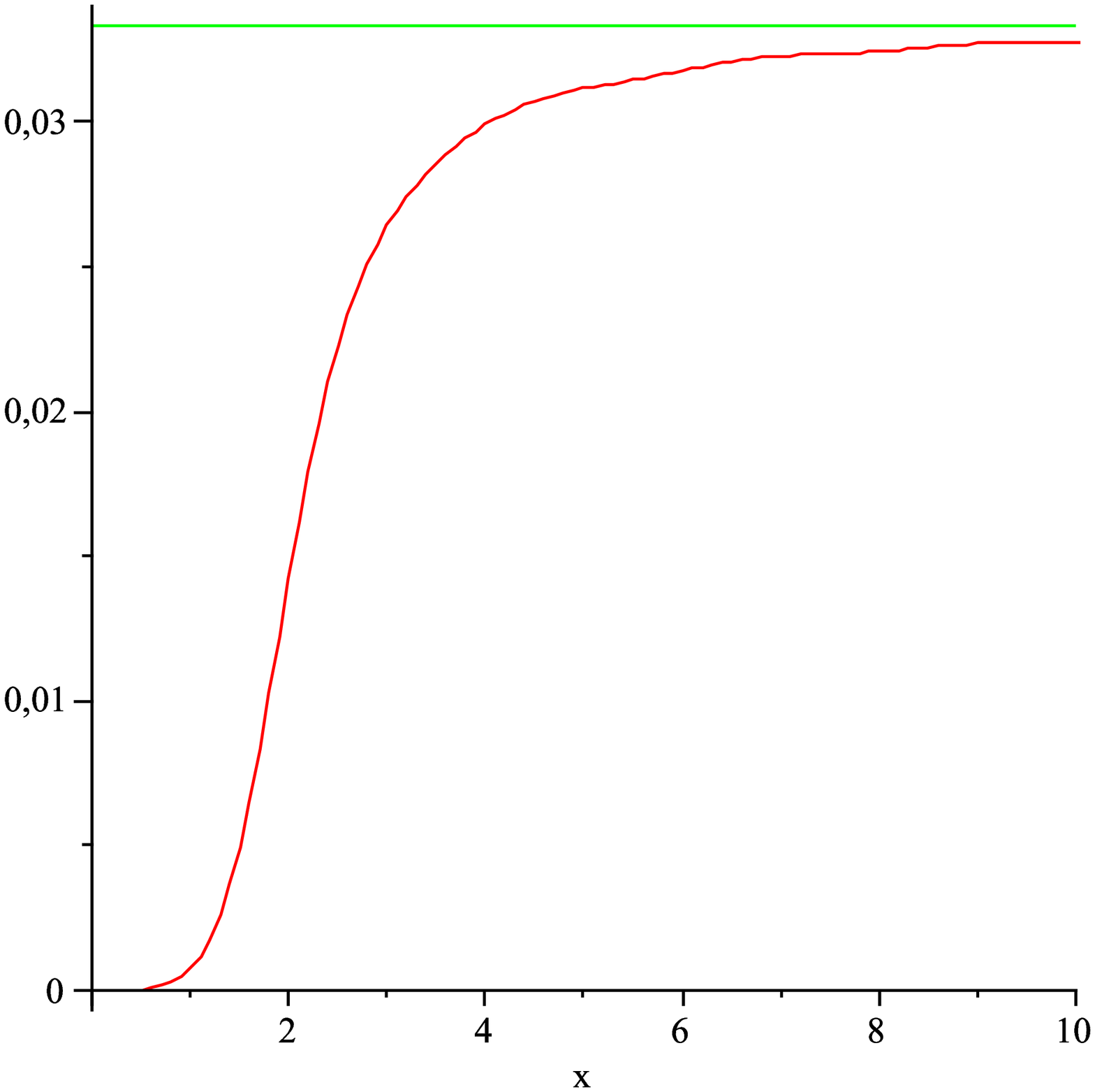}
\\
\includegraphics[{angle=0,width=7cm}]{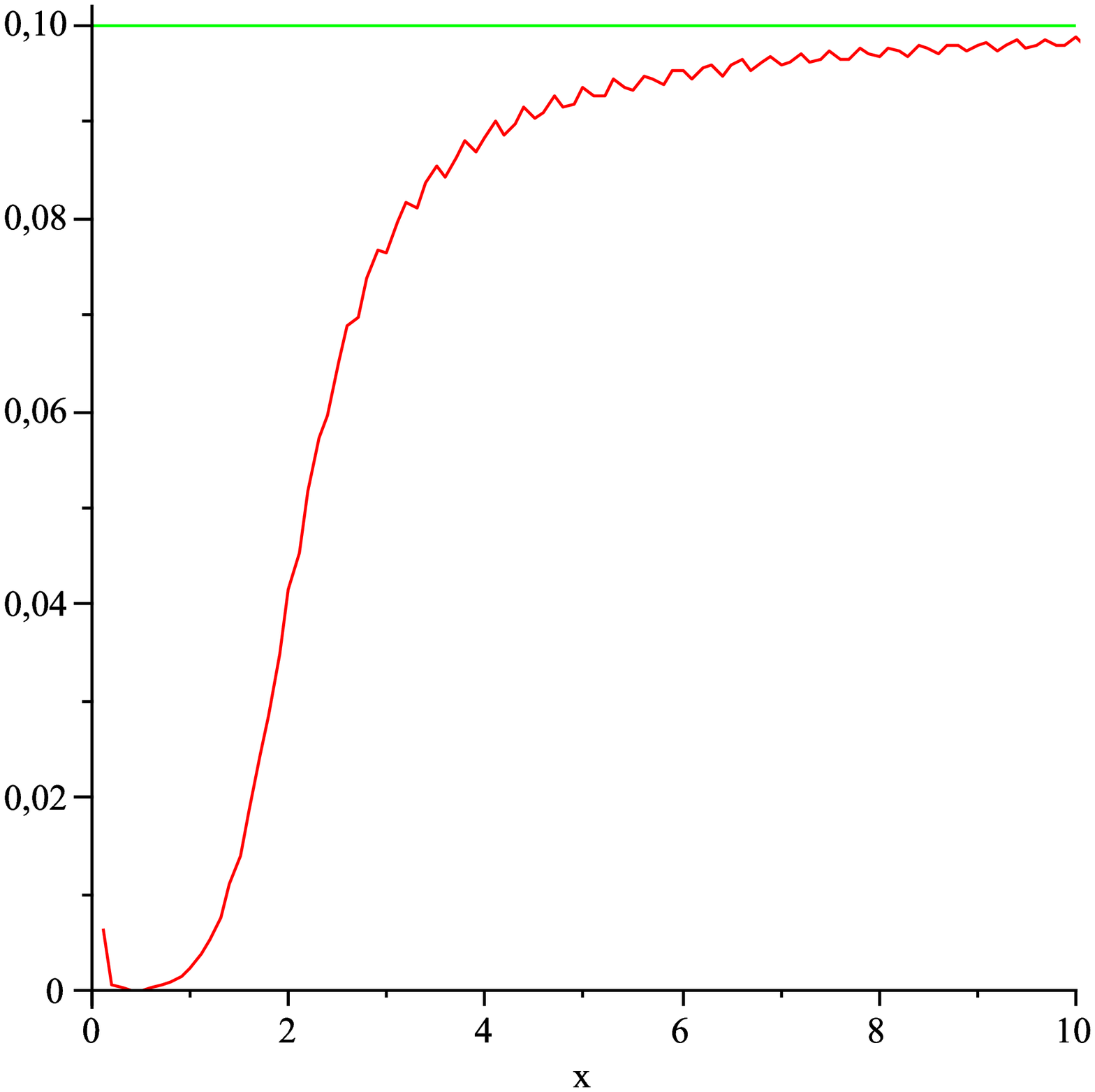}\hspace{0.3cm}
\includegraphics[{angle=0,width=7cm}]{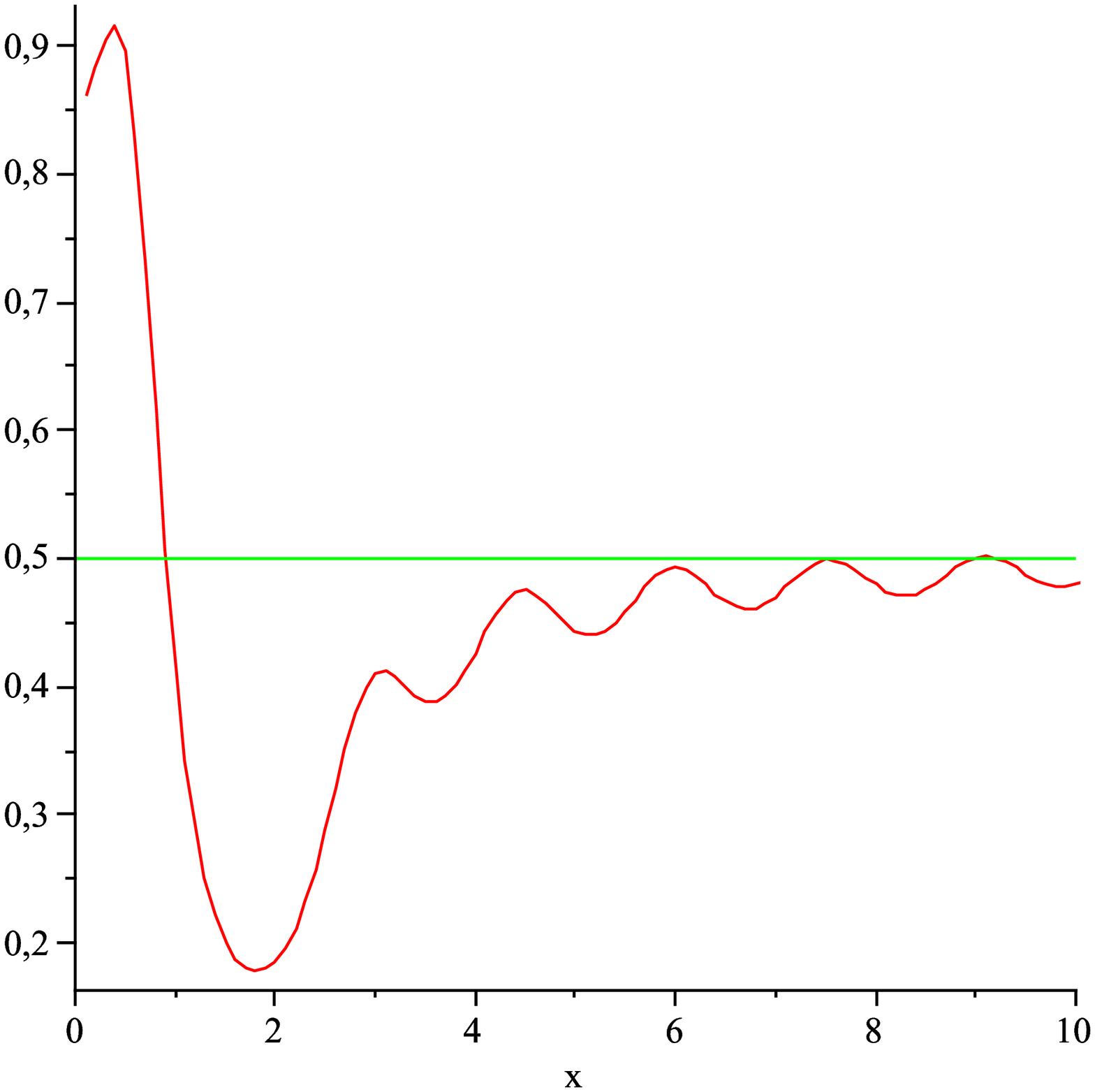}
\caption{Plots of $|\psi_m(0)|^2$ as a function of m. Here each $\psi(z)$ has been solved by Maple with Runge-Kutta default method of
fourth order, taking $z_{max}=30$ (upper), $z_{max}=10$ (lower, left) and $z_{max}=2$ (lower, right). All plots show the asymptotic limit
 corresponding to $1/z_{max}$.}
\label{figprob0}}

We normalize the wavefunctions in a box of size $[-z_{max},
z_{max}]$. However, an important point is to consider a box with
sufficiently large size $z_{max}$ in order for the potential
$V_{sch}$ to achieve the $1/z^2$ regime. In fact, as already noted
in \cite{csaki}, localization of gravity is determined by the far
region of the potential. As we can see in Fig.~\ref{figphiz}, for
the choice of $V_{sch}$ the asymptotic regime is achieved for
$z>~20$. In this way we must choose an at least comparable size
$z_{max}$ for our box in order to correctly reproduce the
gravitational features. However, to illustrate possible mistakes
within the procedure, we will consider boxes of different sizes,
to see spurious effects which can be observed for small boxes. We
plot some results in Fig.~\ref{figprob0}, where we show the
normalized $|\psi_m(0)|^2$ as a function of $m$ for $z_{max}=30$,
$z_{max}=10$ and $z_{max}=2$. Note that for $z_{max}=30$ we
achieved the asymptotic regime for the $1/z^2$ behaviour for the
$V_{sch}$ potential. On the other hand, for smaller values of
$z_{max}$, say $z_{max}=10$, besides a first small peak, small
perturbations resembling resonances start to appear. This
behaviour is greatly reinforced for $z_{max}=2$. Clearly there are
no resonances at all in this problem, the effect just observed
being a consequence of a wrong choice of the size of the box,
which could not accommodate the correct behavior of the potential
$V_{sch}.$

Let us now turn to asymptotic analysis in order to evaluate the
degree of accuracy of our Fig. \ref{figprob0}, for $z_{max}=30$.
First of all, note that in this region the factor $|\psi_m(0)|^2$
increases monotonically with $m,$ achieving a plateu for larger
masses. The very existence of this plateau is confirmed after
considering $m^2>>V_{sch}$ in the Schroedinger equation,
Eq.~(\ref{se}). Indeed, in this case it reduces to \ben
-\frac{d^2\psi_m(z)}{dz^2}=m^2\,\psi_m(z), \,\,m^2>>V_{sch} \een
with solution $\psi_m(z)=N\cos(mz)$. After normalization in a box
one obtains $|\psi_m(0)|^2=1/z_{max}$ for large $m$, which for
$z_{max}=30$ gives $\psi_m(z)\sim1/30$. See Fig.~\ref{figprob0}
for the accuracy of this limit.

Now let us consider the regime of lower modes. According to Ref.~\cite{csaki} -- see particularly Sec. 4 --, this regime is related to
the $z>>1$ behaviour of $V_{sch}$. Indeed, in our case we have the limit $V_{sch}\sim 12/z^2$. This has the particular form
$\alpha(\alpha+1)/z^2,$ already proposed in Ref.~\cite{csaki}. There, it is shown that when this occurs one expects the lower massive
modes to obey the relation $\psi_m(0)\sim m^{\alpha-1}$. For our potential we have $\alpha=3,$ and this leads to $|\psi_m(0)|^2\sim m^{4}$.
In order to look for this power law behavior in $|\psi_m(0)|^2,$ we plot in Fig.~{\ref{figlogprob0}} the former results of Fig.~\ref{figprob0}
for $z_{max}=30$ in a logarithmic scale, focusing on the lower modes.

\FIGURE{
\includegraphics[{angle=0,width=7cm}]{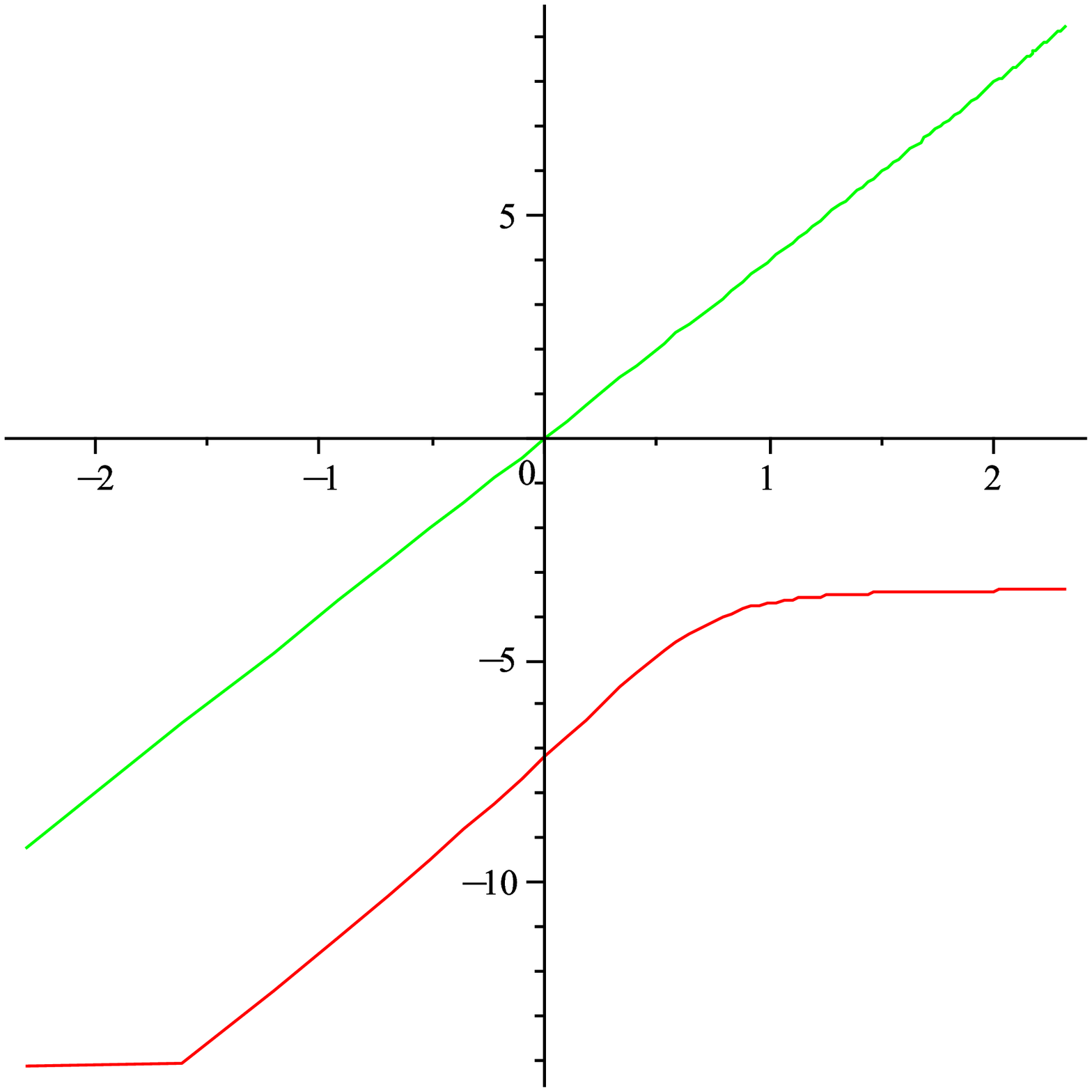}
\includegraphics[{angle=0,width=7cm}]{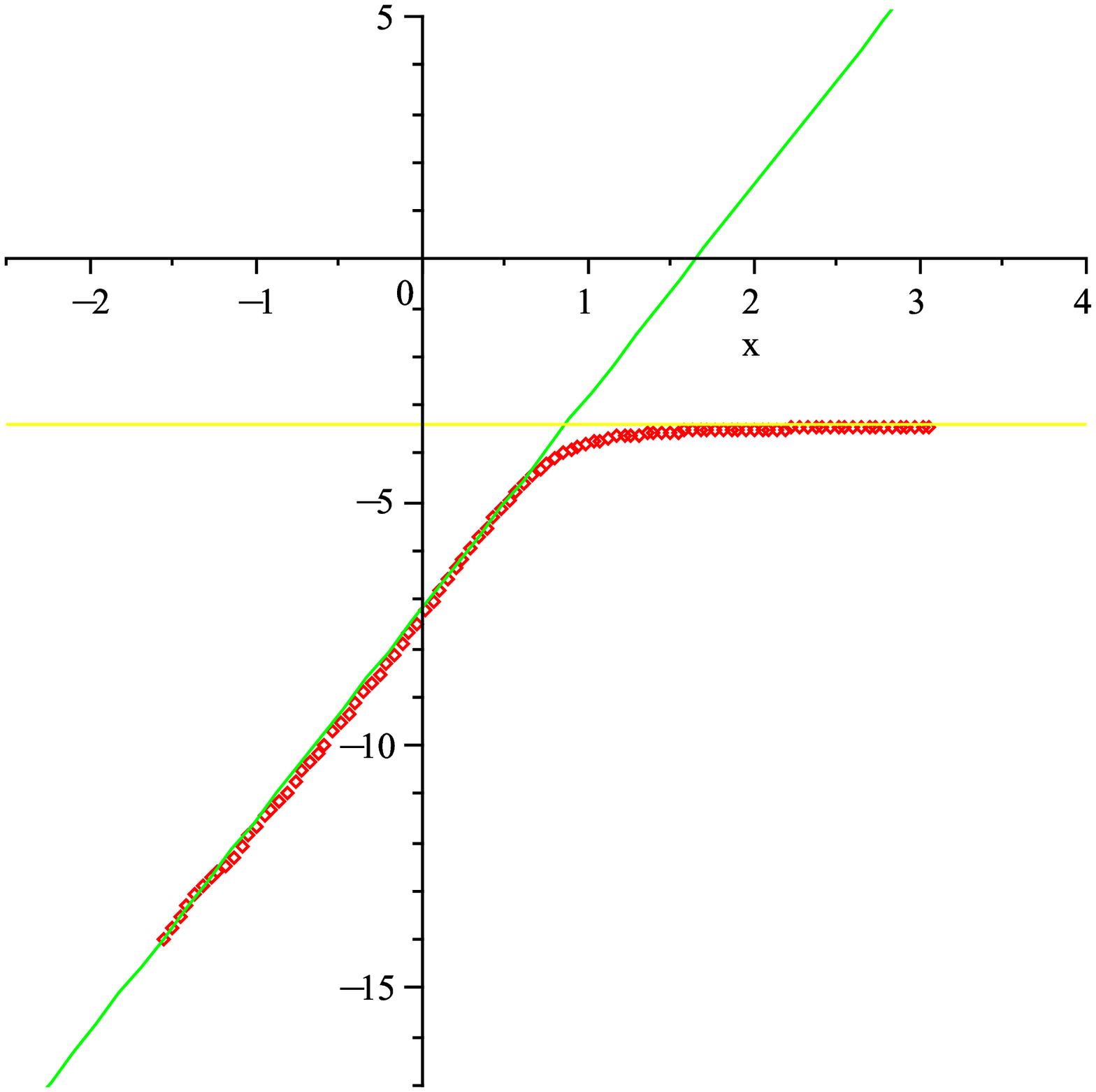}
\caption{Plot of $\log(|\psi_m(0)|^2)$ as a function of $\log(m)$ (left panel). Each $\psi(z)$ was generated with 500 points, taking $z_{max}=30$.
We  are also displaying the straight line corresponding to the limit $|\psi_m(0)|^2\sim m^{4}$. The right panel shows the same plot,
after dropping out the data which come from very small values of $m,$ since they come from the plane wave normalization in a box with size not
large enough. We  are also displaying the straight lines corresponding to the lower (green line) and higher (yellow line) masses.}
\label{figlogprob0}}

Note that the desired behaviour for small masses is obeyed in the model for almost two orders of magnitude. However, for very small masses,
say for $m<m^*$, the approximation breaks down, indicating the formation of a first small peak already noted in Fig.~\ref{figprob0}.
Our simulations show that the region of validity of the approximation $|\psi_m(0)|^2\sim m^{4}$ is enlarged toward smaller masses when the
parameter $z_{max}$ is increased. Qualitatively, this can be understood very easily, since the wavefunctions have larger wavelenghts for
smaller masses. As a consequence, the wavefunctions in this region do not oscillate sufficiently in the box and the procedure of normalization
cannot be justified. With this reasoning, the procedure of carrying out the integration of massive modes to obtain the Newtonian potential
must be done with great care, in order to eliminate the wrong contribution of modes for $0<m<m^*.$ One possible solution, justifiable by our
asymptotic analysis, is to extrapolate the observed polynomial behaviour toward the region $0<m<m^*$, as shown in the right panel of
Fig.~\ref{figlogprob0}.

\subsection{Gravity localization: numerical procedure and asymptotic analysis}

Let us now investigate the Newtonian potential for two masses $m_1$ and $m_2$ separated by a distance $R$.
In order to obtain the potential one sums the tower of Kaluza-Klein excitations to the usual contribution coming from the zero-mode as
\cite{csaki}
\ben\label{UR}
U(R)=G \frac{m_1m_2}{R} + \frac {m_1m_2}{M_*^3}\int_0^{\infty}{dm \frac{e^{-mR}}R |\psi_m(0)|^2},
\een
where $G=M_4^{-2}$ represents the four dimensional coupling, $M_*$ is the fundamental five dimensional
Planck scale and the integration is done considering the brane position at $z=0$, as it is done in the thin brane case. Separating the
 contributions of the Einstein-Hilbert action due to the four dimensional part and the extra dimension, we are led to an expression
  relating the two scales as \cite{csaki}
\ben
M_4^2=M_*^3N_0^2,
\label{eqMM}
\een
where $N_0^2=\frac8{3\pi}$, after the normalization procedure given by Eq.~(\ref{normproc}) is applied to the zero-mode Eq.~(\ref{zeromode}).
 In this way, Eq. (\ref{UR}) can be written as
\ben
U(R)=G \frac{m_1m_2}{R} \biggl[1 + \frac {3\pi}{8}\int_0^{\infty}{dm {e^{-mR}}|\psi_m(0)|^2} \biggr],
\een

Note that in order to obtain the Newtonian potential $U$ for a given separation $R,$ it is sufficient for practical purposes that the
integration goes over the massive modes until masses of $O(10/R)$. For higher distances $R,$ a considerable simplification is achieved, but
for lower distances one must consider higher massive modes. In the limit, for $R\to0$ we must integrate all possible masses until $m\to\infty$.
In this way, another extrapolation of the obtained data for $|\psi_m(0)|^2$ is necessary for higher modes. Namely, one needs to extrapolate
the results displayed in Fig. \ref{figlogprob0}, right panel, as $|\psi_m(0)|^2\simeq1/z_{max}$.

Let us now deal with gravity localization from the asymptotic analysis of $|\psi_m(0)|^2$. We can separate the regime
$|\psi_m(0)|^2=C_1 m^{2(\alpha-1)}$, valid for smaller masses from the re\-gi\-me $|\psi_m(0)|^2=C_2/z_{max}$, valid for larger masses.
The constants $C_1$ and $C_2$ can be learned from Fig. \ref{figlogprob0}, right panel. This figure shows that the transition region can be
located at $\log(m_T)\sim 1$. In this way we can separate both contributions for the Newtonian potential as follows
\ben
\label{URint}
U(R)\simeq G \frac{m_1m_2}{R} \biggl[1 + \frac {3\pi}{8}C_1\int_0^{m_T}{dm {e^{-mR}}m^{2(\alpha-1)}}
+ \frac {3\pi}{8}\int_{m_T}^{\infty}{dm {e^{-mR}}\frac{1}{z_{max}}} \biggr].
\een

For the model under investigation we have $\alpha=3,$ and this leads to
\ben
U(R)&\simeq& G \frac{m_1m_2}{R} \biggl[1-\frac{3\pi}8\frac{C_1}{R^5}\biggl(-24+e^{-m_TR}m_T^4R^4+4m_T^3R^3e^{-m_TR}\nonumber
\\
&+&12m_T^2R^2e^{-m_TR}+24m_TRe^{-m_TR}+24e^{-m_TR} \biggr) + \frac{3\pi}8 \frac{1}{z_{max}}\frac{e^{-m_TR}}R\biggr],
\een
and for $m_T R>>1$ we can write
\ben
U(R)\simeq G \frac{m_1m_2}{R} \biggl[1 + \frac {3\pi}{8}C_1 \frac {24}{R^5} \biggr],
\een
which reproduces the Newtonian potential for large values of R. We then learn from this expression that the present model reproduces the
known gravitational limit at large distances, localizing gravity with a $1/R^6$ correction.

However, the full integration of Eq. (\ref{URint}) is still
necessary in order to estimate the accuracy of our procedure. This
is important in case the potential $V_{sch}$ is not analytically
know and also to find the value of the parameter $C_1$. To
implement this, we fit numerically the logarithmic plot of
Fig.~\ref{figlogprob0}, right panel; we did that considering that
it comes approximately as a superposition of two straight lines,
one for each region where $m< m_T$ and $m>m_T$. We remark that
this is important in order to better achieve a control for models
where the Schroedinger potential are not known analytically. First
of all we get from a linear fitting of the Fig.~\ref{figlogprob0},
right panel, the result $\log |\psi_m(0)|^2=-7.186+4.359\log m $,
leading to $|\psi_m(0)|^2=7.57\times 10^{-4}m^{4.359}$, giving
$C_1=7.57\times 10^{-4}$. From the same figure we can estimate
$m_T$ as the value of $m$ corresponding to the crossing of both
straight lines for larger and smaller masses. This gives
$\log(m_T)=0.868$, or $m_T\simeq2.383$. Also, we find for the
region $m>m_T$ the limit $\log |\psi_m(0)|^2=1/z_{max}\sim1/30$.
After substituting these results in Eq. (\ref{URint}) and
performing the integration, we can find the correction coefficient
$L$ for $U-U_{Newton}\sim1/R^L$, which is displayed in
Fig.~\ref{figseqLlogR}. The numerical results give a correction
$L\simeq6.34$ for large distances, which is to be compared with
$L=6,$ as the value which comes from the asymptotic analysis.

\FIGURE[h]{
\includegraphics[{angle=0,width=7cm}]{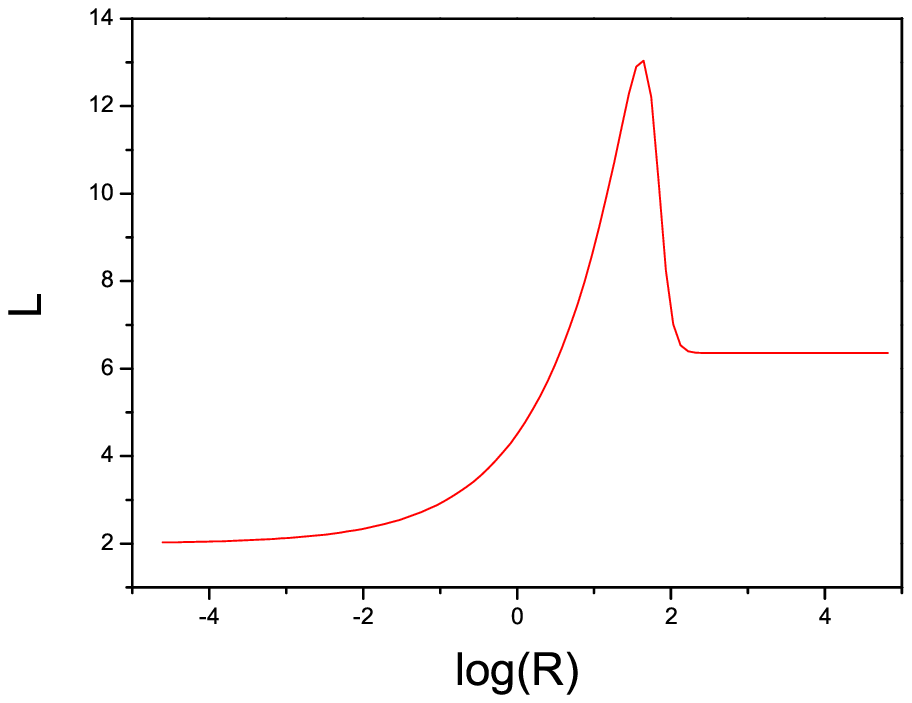}
\caption{Plot of the $L$ coefficient as a function of the distance
$R$ between two unit test masses. The results are obtained after
performing an expansion valid for $R>>1/m_T$ with
$m_T\simeq2.38$.} \label{figseqLlogR}}

\subsection{The Numerov method}

We can solve the same problem for the massive modes using the Numerov method \cite{N} instead of the Runge-Kutta fourth-order method that we have
just used. The Numerov method -- see Appendix -- is largely used in order to look for bound states in quantum mechanics. We can also apply
it here, and some results for this model are displayed in Fig.~\ref{figNumerov}. Despise the slow convergence of the Numerov method compared
with Runge-Kutta methods, we can use the Numerov method when the Schroedinger potential is only known numerically.

The right panel in Fig. \ref{figNumerov} shows a good visual concordance between both methods for $|\psi_m(0)|^2$ as a function of m. However,
in the former subsection we have emphasized the importance of the lower massive modes for gravity localization. A careful analysis in a logarithmic
scale has shown that for small masses, the discrepancy between the methods grows, with Numerov method tending to present
$|\psi_m(0)|^2\sim m^\beta$ with a higher power law $\beta$ than the Runge-Kutta method. Naturally, this may reflect on predicting a
correction to the Newtonian potential with a higher power law.

We will use this method in Sec.~\ref{secnumeric} for some other models, where the Schroedinger-like potential is only known numerically, since
 there the Runge-Kutta method cannot be used anymore.

\FIGURE{
 \label{figNumerov}
\includegraphics[{angle=0,width=6cm}]{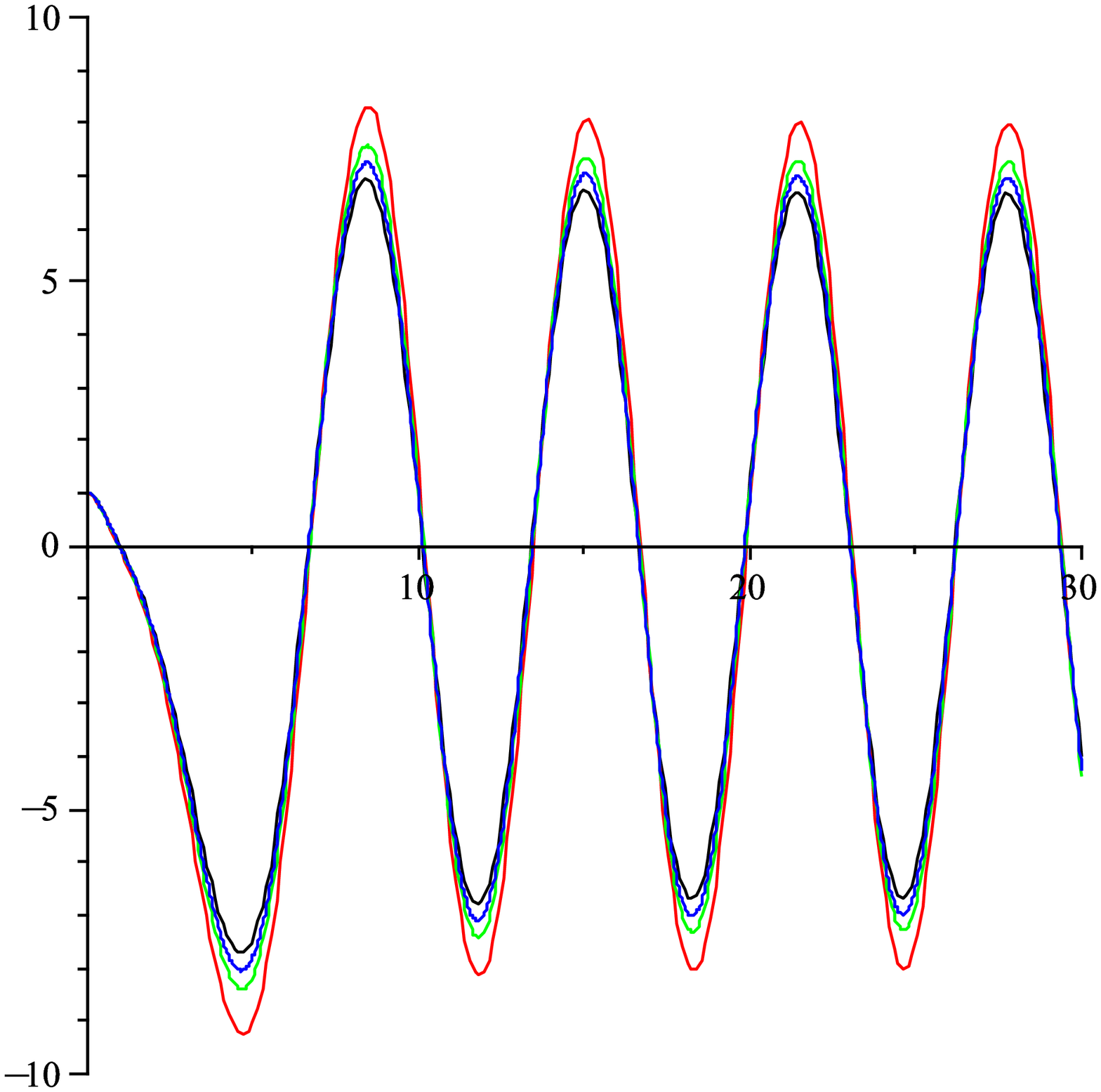}
\includegraphics[{angle=0,width=6cm}]{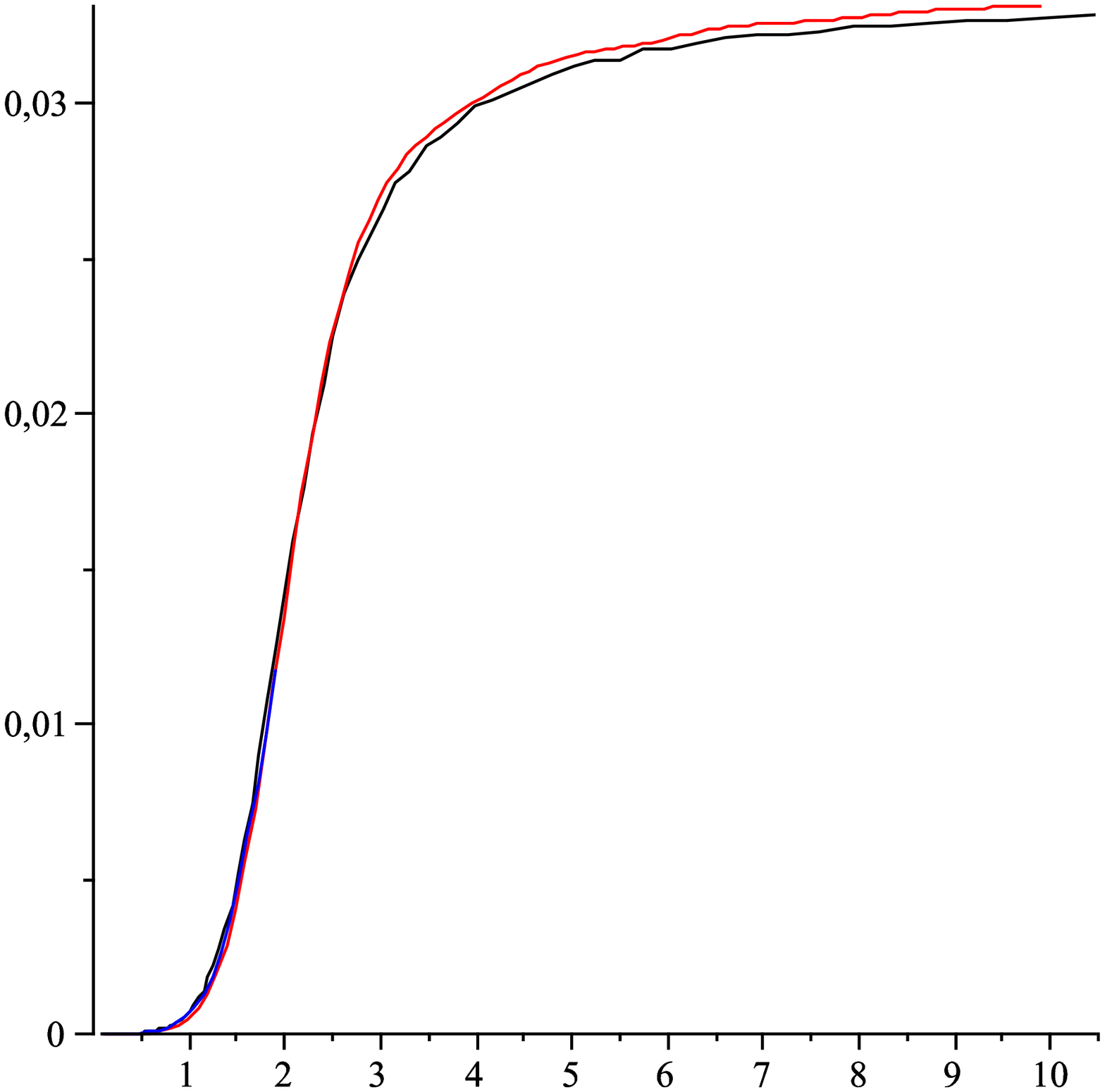}
\caption{Plots of $\psi(z)$ for $m=1$ and $z_{max}=30$ (left). The
lines are displayed to achieve a better comparison between the
default method from Maple (500 points-black line), for the
analytically known $V_{sch}$ and the Numerov method (212-red,
490-green and 980 points-blue lines) applied for $z>0$, better
indicated for numerically known potentials $V_{sch}$.  Note the
slow convergence of the method.  The right panel shows
$|\psi_m(0)|^2$ as a function of m, comparing Runge-Kutta method
(black) with Numerov method for $\psi(z)$ with 212 points ($\delta
z=0.14$, red line) and 980 points ($\delta z=0.0316$, blue
line).}}}

\section{Models with analytic $V_{sch}$: asymptotic analysis}
\label{secanalit}

If we compare with the presented numerical procedure, the asymptotic analysis \cite{csaki} has very good precision to estimate the power
of the first correction to the Newtonian potential. However, the study can only be done when one knows an analytical relation for $y=y(z).$ To
make this point clearer, let us illustrate this possibility with the following examples, recently considered in Ref.~\cite{BBL}.

\subsection{The case $W_1(\phi)=3a\sin(b\phi)$}

For this model we have \ben
A(y)=-2\frac2{3b^2}\log(q\cosh(\frac32ab^2y)). \een In general
this model does not give an analytic relation for $z(y).$ However,
for $b^2=2/3$ we get \ben z=\int e^{-A(y)}dy = \frac qa\sinh(ay),
\een with \ben A(z)=-\log(q\sqrt{\frac{z^2a^2}{q^2}-1}). \een This
model gives an analytic expression for $V_{sch}$, depending on two
parameters $q$ and $a$: \ben
V_{sch}=3a^2\frac{(5z^2a^2+2q^2)}{4(q^2-a^2z^2)^2}. \een In this
way, we can apply the asymptotic analysis in order to study the
Newtonian potential. In fact, we have the limit \ben
V_{sch}=\frac{15}{4z^2},\;\;\;\;\; z>>1. \een We see from the
former expression the independence of the potential of the
parameters $q$ and $a$ for large values of $z$. Following the
former section we can write $V_{sch}={\alpha(\alpha+1)}/{z^2}$ for
$z>>1.$ Thus, $\alpha=3/2$ and we expect that $|\psi_m(0)|^2\sim
m$ for small masses. This gives the same correction $1/R^3$ for
the Newtonian potential as in the Randall-Sundrum model.

\subsection{The case $W_2(\phi)=3a\sinh(b\phi)$}

For this model we have
\ben
A(y)=-\frac{1}{3b^2}\log(\sec^2(\frac{3ab^2y}2)).
\een
For $b^2=1/3$ we can obtain an analytic expression for $z(y)$, namely
\ben
z=\frac2a\tan(\frac{ay}2).
\een
Inverting this expression and substituting in the expression for $A(y)$ we get
\ben
A(z)=-\log\bigl(1+\frac{z^2a^2}4\bigr).
\een
For this model we have an analytic expression for the Schroedinger potential
\ben
V_{sch}=\frac{12a^2(z^2a^2-1)}{(z^2a^2+4)^2}.
\een
with the limit
\ben
V_{sch}=\frac{12}{z^2},\;\;\;\;\;z>>1.
\een
This limit leads to the same asymptotic result of the former section, where $\alpha=3$ and $|\psi_m(0)|^2\sim m^4$, with a $1/R^6$ correction
to the Newtonian potential.

\section{The case of models with numerical $V_{sch}$}
\label{secnumeric}

For a large class of models there is no analytic solution for $y=y(z)$.  In this way, the analysis of $A(z)$ and the potential $V_{sch}$
 must be done numerically with no {\it{a priori}} asymptotic analysis available to guide us. In this case the method presented in this work may
 be used to estimate the behavior of $|\psi_m(0)|^2$ for small masses and to find the first correction to the Newtonian potential. We illustrate
 the procedure with some examples.

\subsection{The case $W_3(\phi)=2a\tan^{-1}(\sinh b\phi)$}

This model gives \cite{BBL}
\ben
A(y)=\frac1{3b^2}\log(1+a^2b^4y^2)-\frac23ay\tan^{-1}(ab^2y). \een
We choose $a=1$ and write $c=b^2$ to get \ben
A(y)=\frac1{3c}\log(1+c^2y^2)-\frac23y\tan^{-1}(cy). \een and now
we study the influence of $c$ on gravity localization. In this
case, even the relation given by Eq. (\ref{express_zy}) for $z(y)$
is not possible to be obtained analytically. Thus, we have
integrated it numerically for some values of c, leading to a
numerical determination of $A(z)$ which we show in
Fig.~\ref{figmodel4}. Gravity localization is confirmed after
demonstrating normalization of the zero-mode, as we also show in
Fig.~\ref{figmodel4}. We have repeated the same analysis in a
logarithmic scale, and the results shown that the zero mode is
nicely normalized as it goes to zero faster than $z^{-1/2}$, thus
confirming the occurrence of gravity localization.

\FIGURE{
\includegraphics[{angle=0,width=6cm}]{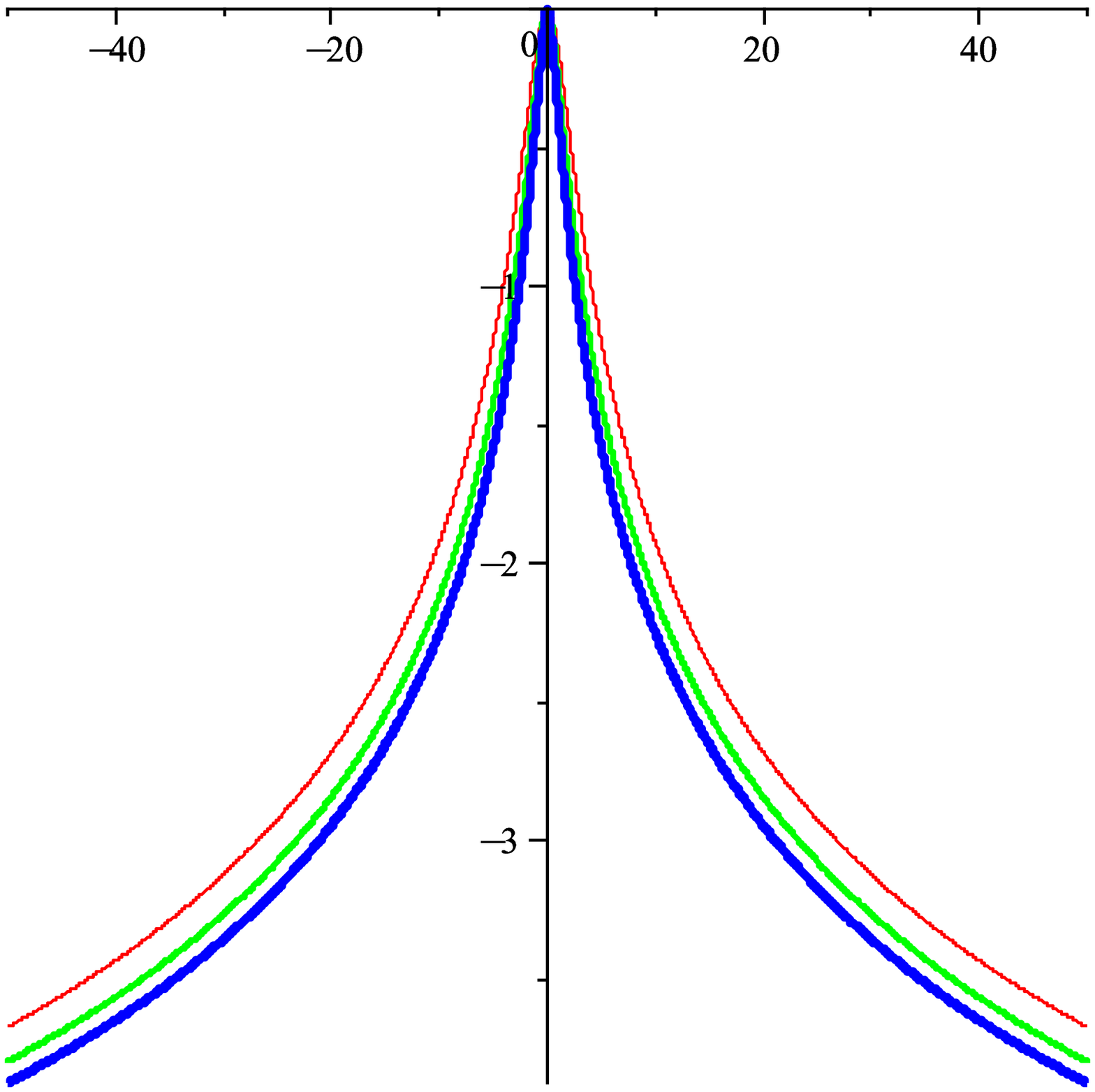}
\includegraphics[{angle=0,width=6cm}]{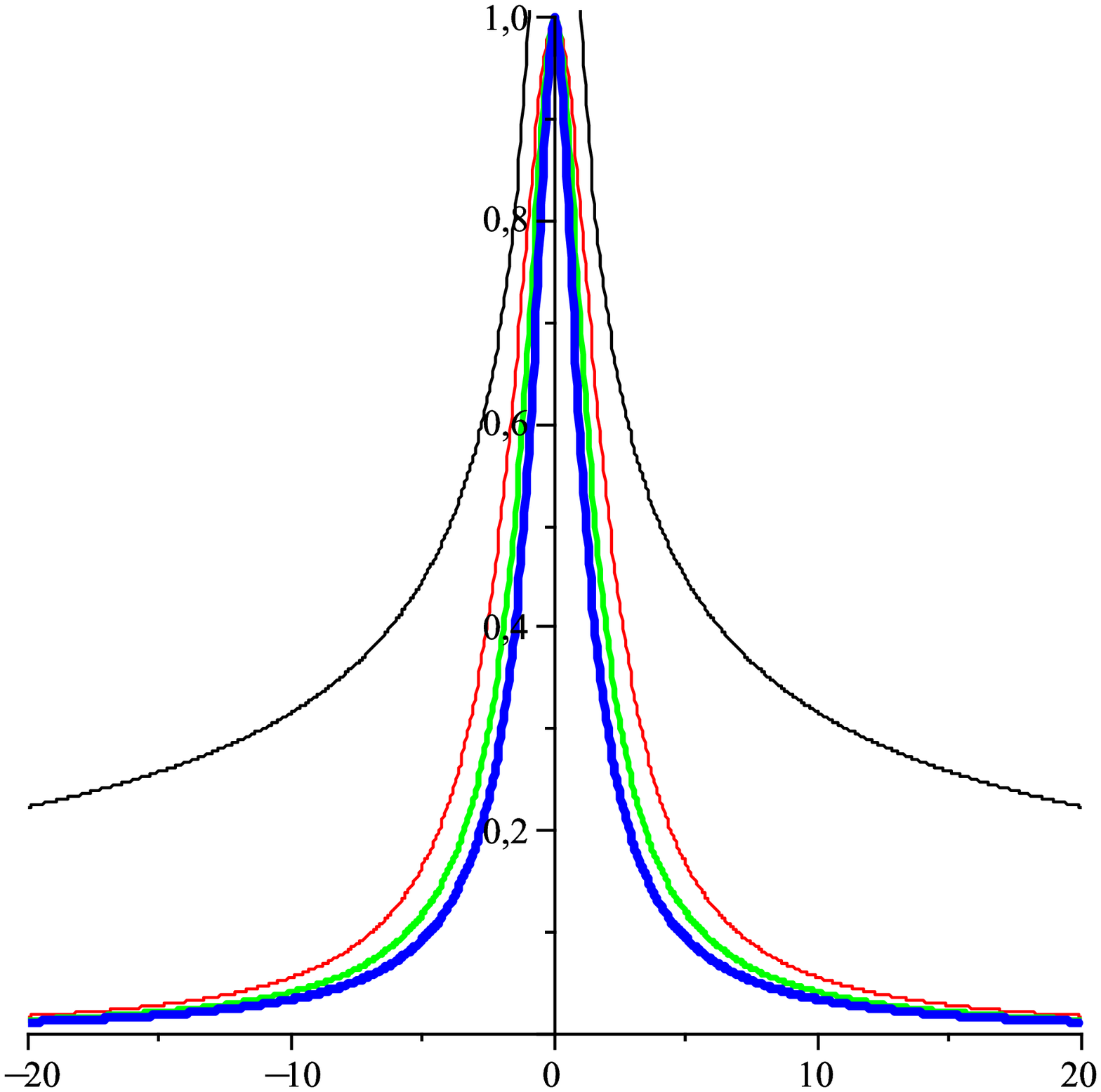}
\caption{Plots of $A(z)$ (left) for the model with $W_4(\phi)=2\tan^{-1}(\sinh b\phi),$ for the values $c=0.5$ (red, thin), $1$ (green, thick)
and $2$ (blue, thicker), with $\delta z=0.05$. We also show the zero-mode $\psi_0(z)\sim e^{3A/2}$ (right) as a function of $z$ for the same
values of $c$, compared with the $z^{-1/2}$ curve (black).}
\label{figmodel4}}

In order to study details of gravity localization one must construct the potential $V_{sch}$. We did this in Fig.~\ref{figgravitymodel4},
extending the $z$ axis in order to guarantee that the asymptotic regime is correctly achieved and that the higher massive modes can be properly
normalized. Fig.~\ref{figgravitymodel4} shows that even without an analytic expression for $V_{sch},$ it indicates that asymptotically
one has $V_{sch}=\alpha(\alpha+1)/z^2$, and that this regime is better achieved for larger values of $c$. In particular, for the case $c=0.5$ we
 have found for the limit of $V_{sch}z^2$ the  value $4.02(2)$ which gives $\alpha=1.52(4)$.

\FIGURE{\includegraphics[{angle=0,width=6cm}]{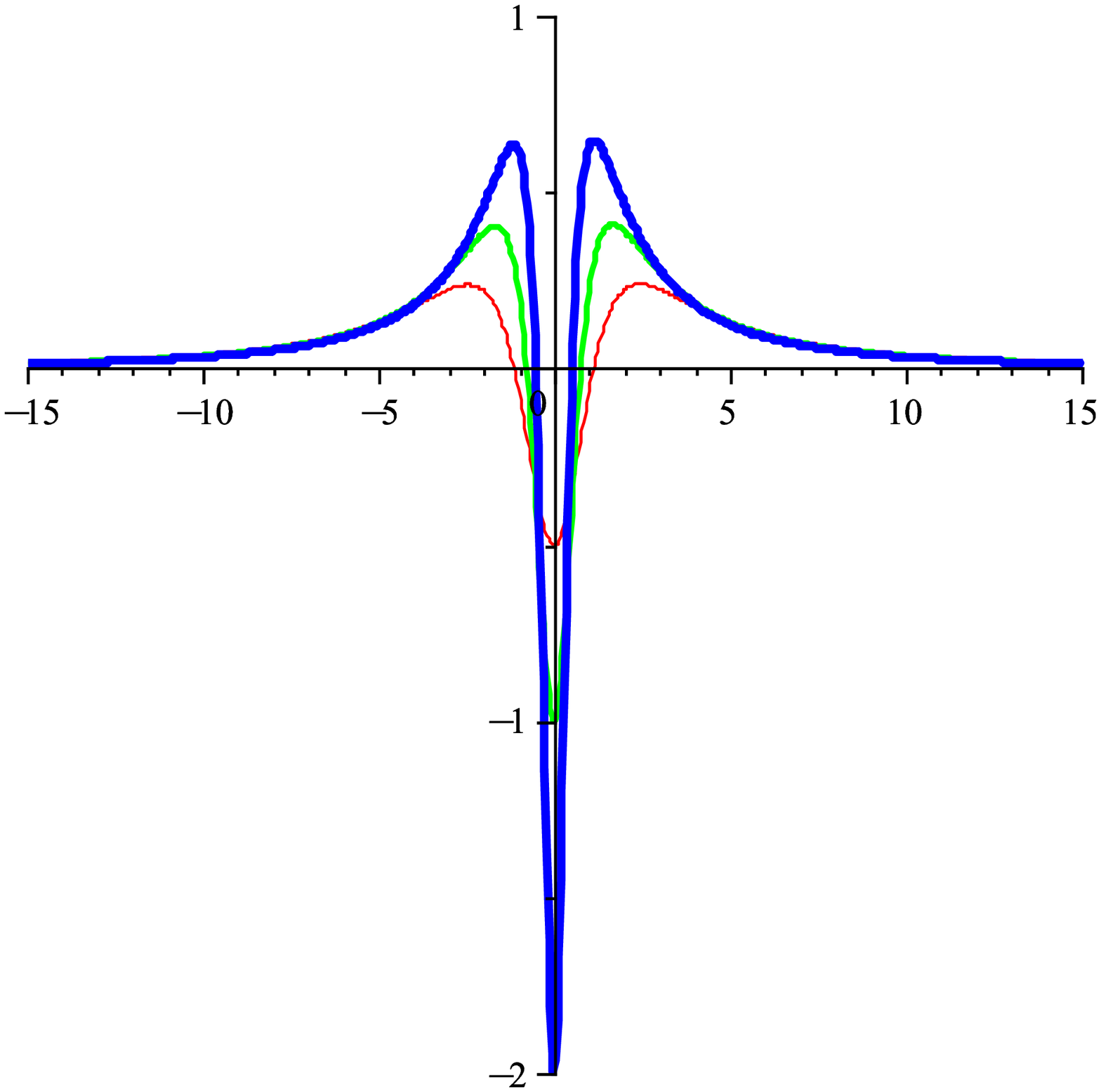}
\includegraphics[{angle=0,width=6cm}]{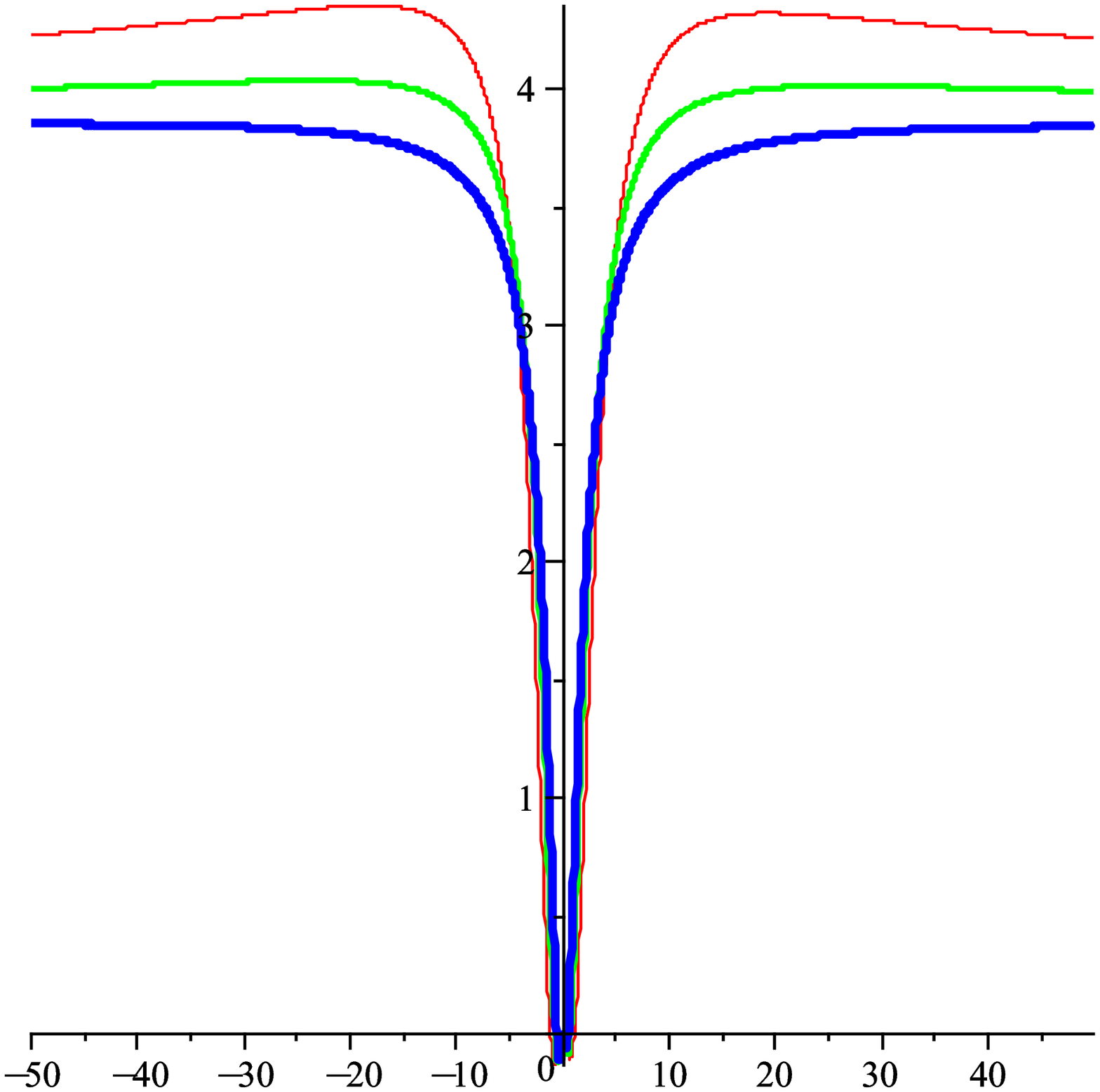}
\caption{Plots of $V_{sch}(z)$ (left) for $W_4(\phi)=2a\tan^{-1}(\sinh b\phi)$ and $z^2V_{sch}(z)$ (right), showing that the asymptotic regime
 is nicely achieved for larger values of $z$. We are using the same conventions of Fig.~9.}
\label{figgravitymodel4}}

In Fig.~\ref{Bfiggravitymodel4}, upper panel,  we show the
normalized probability on the brane $|\psi_m(0)|^2$ as a function
of $m$. We see that for lower values of $c$ (red thin line) the
influence of massive modes increase faster than for higher values
of $c$ (blue thick line). This effect shows that gravity is easier
localized for higher values of $c$, in agreement with the results
shown in Fig. \ref{figmodel4}.

\FIGURE{\hspace{2cm}
\includegraphics[{angle=0,width=6cm}]{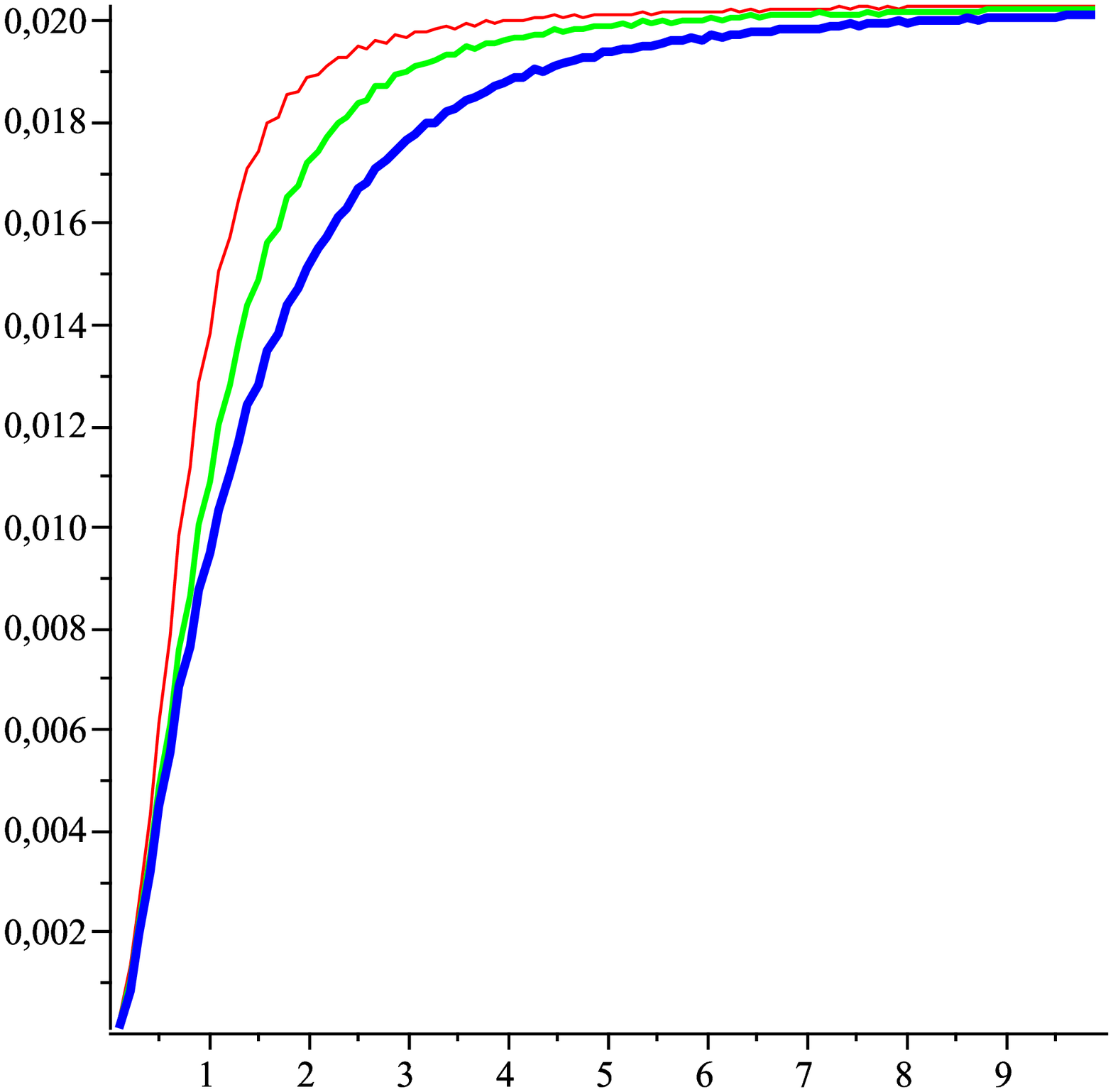}
\\
\includegraphics[{angle=0,width=6cm}]{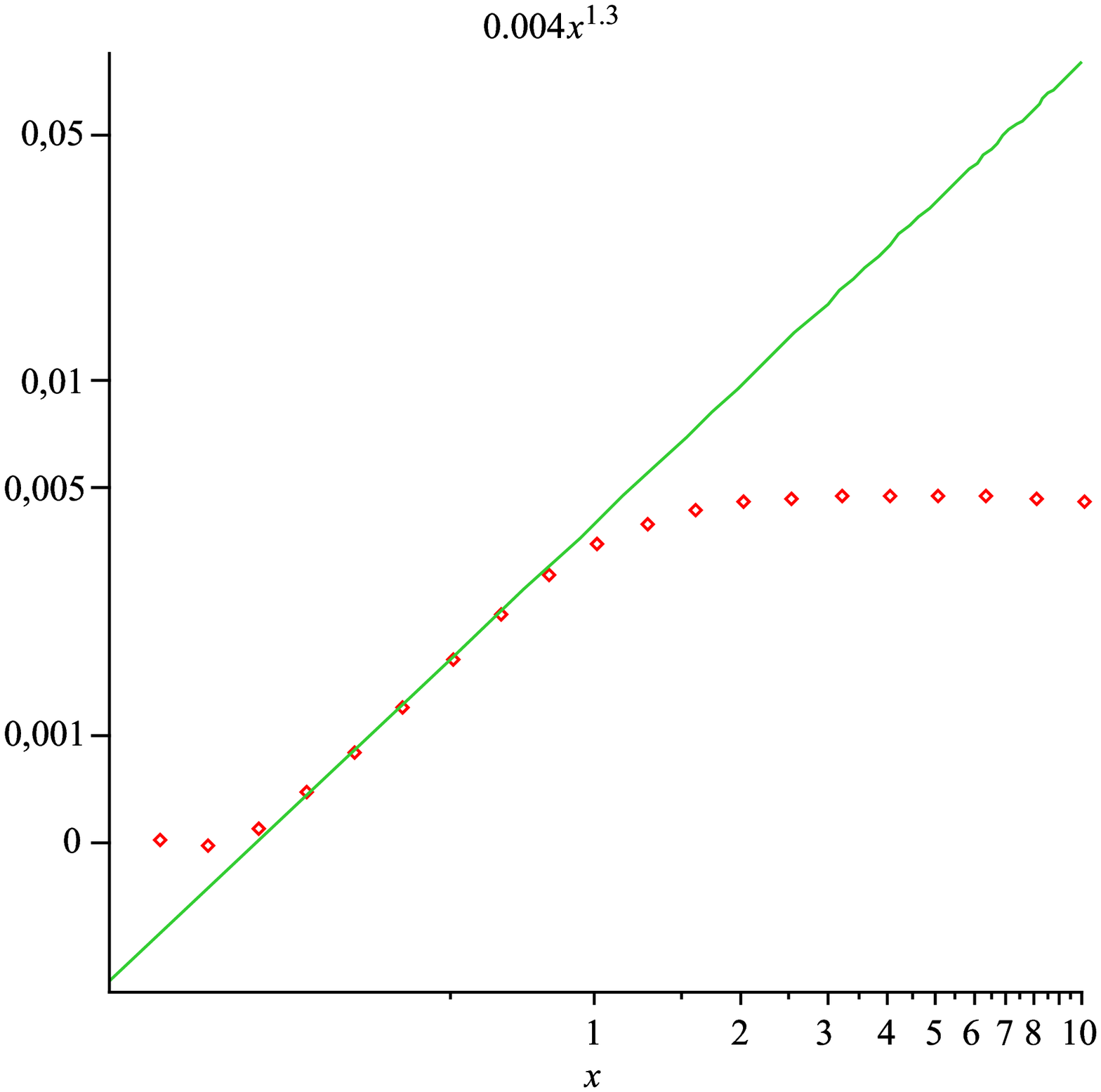}\hspace{2cm}
\includegraphics[{angle=0,width=6cm}]{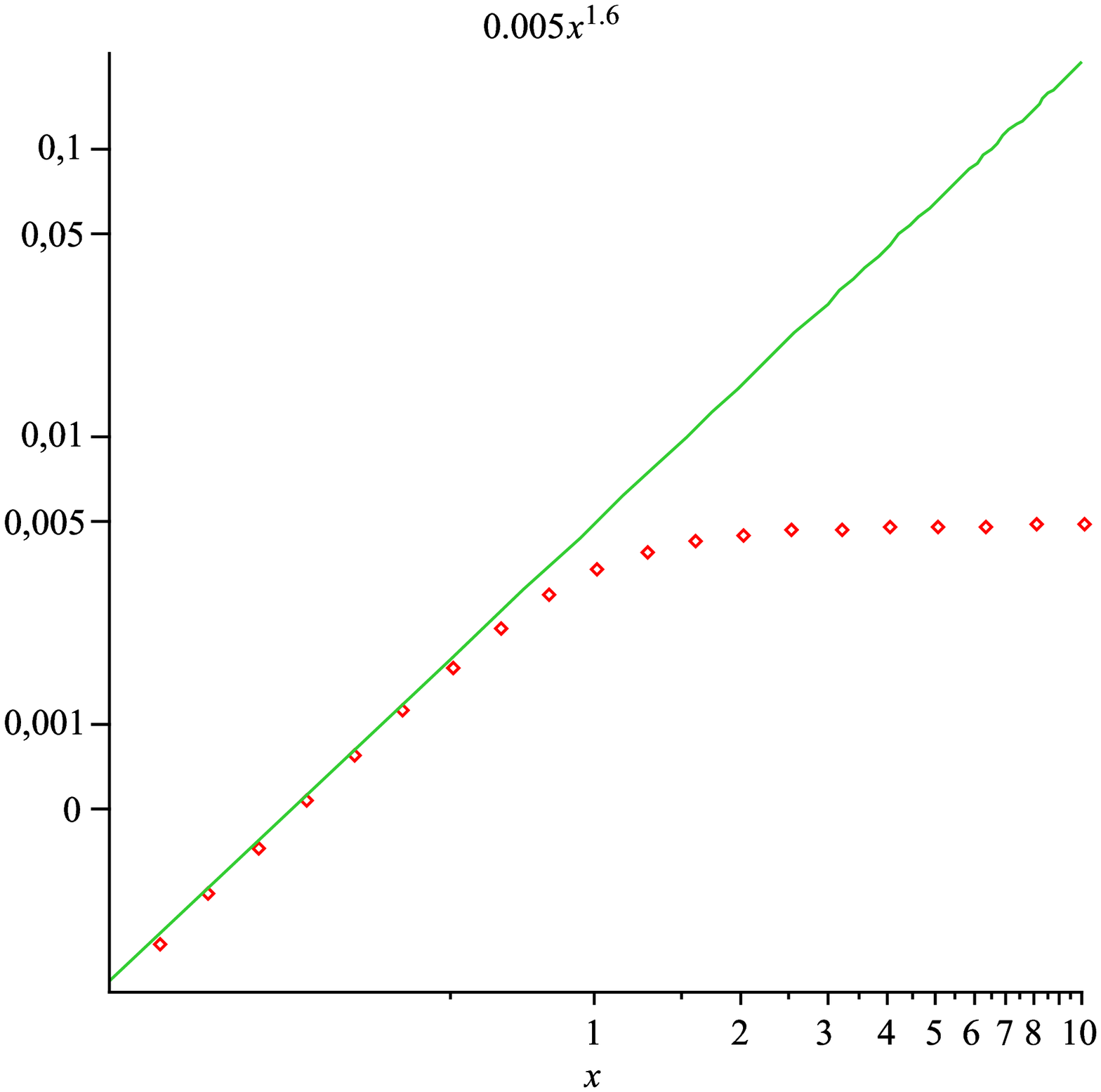}
\caption{Plot of $|\psi_m(0)|^2(m)$ (upper panel), using $y_1=\sqrt{0.02}\simeq0.14$ with $700$ points ($\delta z=0.14$) leading to $y_{max}=4.72,$
using the same conventions of Fig.~9. In the lower panel we plot refined analysis for $c=0.5$, showing logarithmic plots of $|\psi_m(0)|^2(m)$ for
$z_{max}\simeq200$ with $4000$ points, using Euler method (left) and Numerov method (right), fixing the step on $z$ with precision of $10^{-7}$.}
\label{Bfiggravitymodel4}}

Now, the behavior of $V_{sch}(z)$ for $z>>1$ leads to $|\psi_m(0)|^2\sim m^{2(\alpha-1)}\sim m^{1.1(1)}$ for lower masses. We compare this
result with the one shown in Fig.~\ref{Bfiggravitymodel4}, left,  the green line which represents the best fit for lower masses, obtained after
neglecting the very low points. In this figure, we show the logarithmic plot of $|\psi_m(0)|^2$ as a function of $m$, after considering
$z_{max}=200$ with step $\delta z=0.05,$ using the Euler method for determining the massive modes. Note that, as the number of iteractions
increased, it is important to find with higher precision the $y$ points corresponding to $z$ with fixed step. The plot indicates the appearance
of a power law $|\psi_m(0)|^2\sim m^\beta$ for lower masses, with $\beta_{Euler}=1.3$, or after using $\beta=2(\alpha-1)$, $\alpha_{Euler}=1.65$.

The very same simulation with Numerov method gives $\beta_{Numerov}=1.6$ or $\alpha_{Numerov}=1.80$ -- see Fig.~\ref{Bfiggravitymodel4}, right panel.
We notice that the Numerov method produces consistent results both for higher and lower masses, as we would expect from a higher precision method.
On the contrary, the Euler method deviates from expected results in those limits due to error propagation. However, as the simpler Euler method
gives fast results, it was used as a guide to search for the adequate regions to apply the Numerov method. In this way, we found two
complementary numerical estimations for the correction $1/R^L$ for the Newtonian potential for large distances,  $L={\beta+2}$,
leading us to the results $L_{Euler}\sim{3.3}$ and $L_{Numerov}\sim{3.8}$. These results are to be compared to $L_{Schroedinger}\sim{3.1(1)},$
which was produced by asymptotic analysis of the Schroedinger potential.

\subsection{The case $W_4(\phi)=(1/2)m{\phi^2}+b$}

For this model we have
\ben
A(y)=-\frac 16 e^{my} - \frac{by}3.
\een
In general one cannot obtain an analytic expression for $z(y)$. However, for $b=0$ we can use the exponential integral to write
\ben
z=-\frac1m Ei\big(1,-\frac{e^{my}}6 \big)
\een
but we have been unable to invert this expression in order to get an analytic expression for $y(z)$.

\FIGURE{
\includegraphics[{angle=0,width=7cm}]{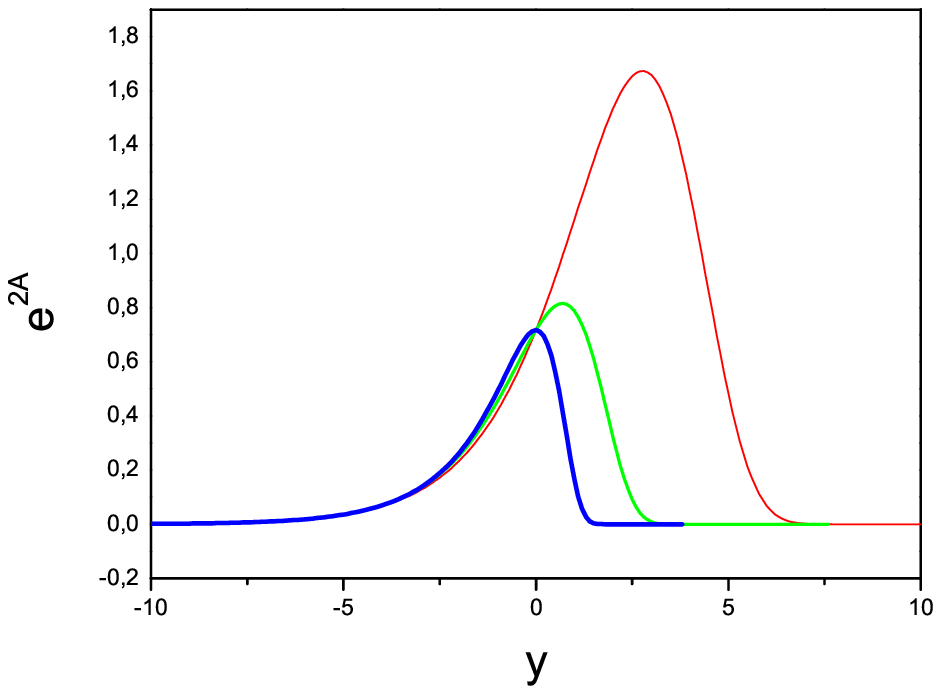}
\caption{Plots of $e^{2A(z)}$ for the $W_3(\phi)$ model, for the values $b=-1$ and $m=0.5$ (red, thin), $m=1$ (green, thick) and $m=2$ (blue, thicker).}
\label{w4_warp}}

In general, this class of models leads to an asymmetric warp
factor. In particular, for negative values of $b$ the warp factor
goes to zero at a negative $z$ value whose module increases for
smaller values of $b$. An illustration of this is given in Fig.
\ref{w4_warp} for $b=-1$, in which one depicts the obtained warp
factor. The figure shows the asymmetry of the constructed brane,
and also the influence the $m$ factor induces on the positive $z$
region, with a maximum warp factor that increases for lower values
of $m$. The case of asymmetric brane has been studied in
\cite{AB}.

The asymmetry of this model poses interesting questions concerning
the influence of the asymmetry on gravity localization, since it
leads to an asymmetric function $z(y)$ as well as an asymmetric
$V_{sch}$. Usually, the massive modes tend to acquire the
characteristics of plane waves far from the brane and the
normalization for symmetric models is realized after considering a
symmetric box with length $L$. For asymmetric models, however, the
normalization procedure for the massive modes must be done with
care and will be postponed to a future investigation.

\section{Conclusions}
\label{secconcl}

In this work we have investigated gravity localization in diverse braneworld models. In general, we can classify the models into
two distinct classes, one which leads to analytic Schroedinger-like potential, and the other, in which the Schroedinger-like potential
cannot be given analytically. The first class of models is very important, and it was nicely studied in Ref.~\cite{csaki}, where an interesting
procedure to fully investigate stability, with a very nice recipe to qualify  gravity localization and Newtonian correction and to quantify
the Newtonian correction.

However, there are interesting models which lead to Schroedinger-like potentials that cannot be given analytically. For such models,
however, the method developed in \cite{csaki} cannot be used to obtain the massive models, thus inducing a gap on the study of gravity
localization in braneworld models. This problem is yet more important within the context of local localization of gravity \cite{KR},
since there the knowledge of the massive modes is crucial to understand the local localization of gravity. This fact has inspired us to
propose the present study, to close the aforementioned gap in the braneworld proposal.

In the course of our investigations, we had to face several interesting issues, among them we would like to pinpoint the following:

\begin{enumerate}

\item{In the Schroedinger-like equation, in order to correctly normalize the eigenfunctions we have to properly choose the size of the
box, the $z_{max}.$ As we have shown, we have to deal with this with care, in order to prevent the inclusion of fake resonant states.}

\item{When the Schroedinger-like potential may be obtained analytically, a comparison between the Runge-Kutta and Numerov methods was done.
In this case, the Runge-Kutta method was applyed directly as a default method for Maple, with an absolute error that can be easily fixed to
a small value. As expected, the best fit between both methods is for Numerov method with smaller steps $\delta z$.}

\item{When the Shroedinger-like potential is analytically known, the asymptotic analysis is the best choice in order to obtain the power law
of the first correction to the Newtonian potential.}

\item{Despide the Numerov method being the best method used to find bound states, its precision is reduced when one changes from a
boundary condition problem to a initial value problem. However, its stability is best seen when compared to Euler method. Also,
the Runge-Kutta method appears to be inadequate when the potential is not known analytically.}

\end{enumerate}

We believe that the present study contributes to enlarge the scope of stability in the braneworld scenario in the presence of large
extra dimensions.
The importance of the numerical procedure widens with the current interest on asymmetric branes, since there the extra dimension drives
the system to behave differently, depending on the positive or negative sense one spans the fifth dimension. We hope to return to this subject
in another work, investigating the presence of massive modes for asymmetric branes.

{\bf Acknowledgements:} The authors would like to thank F.A. Brito and R. Menezes for discussions, and CAPES, CNPq, PADCT-MCT-CNPq,
and PRONEX-CNPq-FAPESQ for financial support.

\section*{Appendix: Construction of eigenfunctions of Schroedinger equation by Numerov method}
\label{ap_numerov}

Consider the Schroedinger equation
\ben \label{ap1_sch} {\hspace{4cm}}\frac{d^2\psi_m(z)}{dz^2} = [V_{sch}(z)-m^2]\psi_m(z){\hspace{4.8cm}} \nonumber(A.1)
\een
Given an energy $m^2$, we look for the corresponding function
$\psi_m(x)$, when the potential $V_{sch}(z)$ is an even function
known only numerically. We follow the theoretical analysis of the
renormalized Numerov method \cite{N}, where the linear character
and the absence of the first derivative on the Schroedinger
equation conspire to achieve a powerful sixth-order numerical
method. Then, defining $f(z)=V_{sch}(z)-m^2$, we can
rewrite Eq.~(A.1) as $\psi''(z)=f(z)\psi(z)$. We consider the
potential $V_{sch}(z)$ given as a set of points, with the variable
$z$ described with fixed step $h.$ This is very important and in
the present case where the relation $y(z)$ -- see Eq.~(\ref{dz})
-- cannot be described analytically one needs a specific numerical
procedure, where a convenient variable step on $y$ leads to a
constant step on $z$. We then expand the functions $\psi(z+h)$ and
$\psi(z-h)$ in Taylor series in order to get
\ben
\label{ap1_taylor} \psi(z+h)+\psi(z-h) = 2\psi(z) + h^2\psi^{(2)}(z) + \frac 1{12}h^4\psi^{(4)}(z) + \frac1{360}h^6\psi^{(6)}(z) + h.o.t.\nonumber{\hspace{0.9cm}}(A.2)
\een
On the other hand, the second derivative of both sides of the
former equation (multiplied by the convenient numerical factor
$-(1/{12})h^2$) leads to
\ben \label{ap1_d2psi}-\frac1{12}h^2[\psi''(z+h)+\psi''(z-h)]=
-\frac16h^2\psi^{(2)}(z)-\frac1{12}h^4\psi^{(4)}-\frac1{144}h^6\psi^{(6)}(z)+h.o.t.\nonumber{\hspace{0.3cm}}(A.3)
\een
Adding both Eqs.(A.2) and (A.3) after substituting (A.1) on
the terms where the second derivative appears, we get
\begin{eqnarray}
&&\psi(z+h)+\psi(z-h)-\nonumber
\\
&& \frac1{12} h^2[f(z+h)\psi(z+h)+f(z-h) \psi(z-h)]=2\psi(z)+\frac 56 h^2f(z)\psi(z)+h.o.t.\nonumber{\hspace{1.2cm}}(A.4)
\end{eqnarray}
We now define $T(z)=(1/12) h^2f(z)$ to obtain the recurrence formula for $\psi(z)$:
\ben\label{ap1_f1recorr}
[1-T(z+h)]\psi(z+h) + [1-T(z-h)]\psi(z-h) = 2[1+5T(z)]\psi(z)\nonumber{\hspace{2.84cm}}(A.5)
\een

We consider that the boundary condition is $\psi(0)=1$ and that we are searching only for even modes $\psi_m(z)$. Then, we apply the recurrence
formula for $z=0$:
\ben
{\hspace{2cm}}[1-T(h)]\psi(h) + [1-T(-h)]\psi(-h) = 2[1+5T(0)]\psi(0)\nonumber{\hspace{2.9cm}}(A.6)
\een
The condition that $V_{sch}(z)$ is even gives $T(-h)=T(h)$. In this way the former equation becomes
\ben
{\hspace{5.2cm}}\psi(h) = \frac{1+5T(0)}{1-T(h)}\nonumber{\hspace{6cm}}(A.7)
\een
This result and the recurrence formula (A.5) lead us to the following strategy: we divide the interval for $z>0$ in $N$ slices, getting the points $z_1,z_2,...z_N$. Defining $\psi(z_1)=1$, we obtain  $\psi(z_2) = (1+5T(z_1))/(1-T(z_2))$. Thus, for  $3\le i\le N$ we construct the points
\ben
{\hspace{2.1cm}}\psi(z_i)=\frac{2[1+5T(z_{i-1})]\psi(z_{i-1})-[1-T(z_{i-2})]\psi(z_{i-2})}{1-T(z_i)}\nonumber{\hspace{3cm}}(A.8)
\een
in order to obtain the desired massive mode as a set of $N$ points, as we use in the text.

\end{document}